\documentclass[preprint,nofootinbib]{revtex4}
\usepackage{graphicx}
\usepackage{epsfig}
\usepackage{amssymb}
\usepackage{color}
\usepackage{amsmath}
\usepackage{lipsum}
\usepackage{graphics}

\newcommand{\affilmark}[1]{\rlap{\textsuperscript{\itshape#1}}}

\newcommand{\CC}{\Lambda}
\newcommand{\rL}{\rho_{\CC}}

\usepackage{epsf}
\usepackage{psfrag}
\usepackage{epstopdf}
\usepackage{graphicx}
\usepackage{caption}
\usepackage{epstopdf}
\usepackage{epsfig}
\begin{document}
\begin{titlepage}

\vspace{0.5cm}

\begin{center}
{\Large \bf Higgs potential from extended Brans-Dicke theory and the time-evolution of the fundamental constants}
\end{center}

%

\begin{center}
\large Joan Sol\`{a}\affilmark{a,b},\quad\
Elahe Karimkhani\affilmark{a,c},\quad
\end{center}
\begin{center}
\normalsize\itshape
\textsuperscript{a}\,Departament de F\'\i sica Qu\`antica i Astrof\'\i sica,\\ and \\ \textsuperscript{b}\, Institute of
Cosmos Sciences (ICCUB),\\  Universitat de Barcelona, Av. Diagonal 647,
 E-08028 Barcelona, Catalonia, Spain\\
 \end{center}
\begin{center}
%
 A. Khodam-Mohammadi\affilmark{}\\[2ex]
\begin{center}
\normalsize\itshape
\textsuperscript{c}\,Department of Physics, Faculty of Science, Bu-Ali Sina
University, Hamedan 65178, Iran\\[1ex]
\end{center}
\end{center}
\vspace*{3ex}

 \vspace{0.2cm}
\centerline{\bf Abstract}
\bigskip
Despite the enormous significance of the Higgs potential in the context of the Standard Model of electroweak interactions and in Grand Unified Theories, its ultimate origin is fundamentally unknown and must be introduced by hand in accordance with the underlying gauge symmetry and the requirement of renormalizability. Here we propose a more physical motivation for the structure of the Higgs potential, which we derive from a generalized Brans-Dicke (BD) theory containing two interacting scalar fields. One of these fields is coupled to curvature as in the BD formulation, whereas the other is coupled to gravity both derivatively and non-derivatively through the curvature scalar and the Ricci tensor. By requiring that the cosmological solutions of the model are consistent with observations, we show that the effective scalar field potential adopts the Higgs potential form with a mildly time-evolving vacuum expectation value. This residual vacuum dynamics could be responsible for the possible time variation of the fundamental constants, and is reminiscent of former Bjorken's ideas on the cosmological constant problem.

\vspace{0.3cm}

\noindent Key words: cosmology, Higgs bosons, cosmological constant, fundamental constants\\
PACS numbers: 98.80.-k, 12.60.Fr, 98.80.Es, 06.20.Jr

\end{titlepage}

\pagestyle{plain} \baselineskip 0.75cm

\section{Introduction}

The finding of the Higgs boson\,\cite{FindingH2012} has obviously been a highly celebrated event in the particle physics world, but at the same time it has triggered a plethora of new questions and posed many other  problems of fundamental nature. Perhaps the most conspicuous one, at least in the particle physics domain, is the hierarchy problem and the associated naturalness problemç\,\cite{HierarchyProblem}, namely the fact that all (fundamental) scalar particle masses are quadratically sensitive to the presence of large scales, typically of order $M_X\sim 10^{16}$ GeV, which are characteristic of Grand Unified Theories (GUT's). As a result, it is not very natural to conceive a relatively light Higgs particle of mass $M_H\simeq 125$ GeV as being part of the standard model (SM) of strong and electroweak (EW) interactions, if the SM is embedded in a GUT. Furthermore, if we look in the cosmological realm, we encounter other  no less unaccountable implications of the Higgs finding, the main one being that it has greatly accentuated the physical (not just the formal) nature of the cosmological constant (CC) problem\,\cite{CCProblem}. Recall that the cosmological constant, $\CC$,  has traditionally been associated to the concept of vacuum energy density, $\rL = \CC/(8\pi G)$, where $G$ is the Newtonian constant, so we should expect that the size of the EW vacuum is as ``real'' or ``unreal'' as the found Higgs boson itself, one cannot exist without the other. In other words, the fact that the EW vacuum energy density is now proven to be $\sim 10^{56}$ times bigger than the measured value of $\rL\sim 10^{-47}$GeV$^4$ does create a phenomenal conundrum for us to elucidate! \cite{CCProblem} -- see also \,\cite{JSPReview2013,GRF2015,SolaGomez2015,JSPReview2016} for additional discussions on the CC problem. The severe difficulties inherent both in the particle physics and in the gravitational domains do indeed crave for new insights as to the very origin of the Higgs potential, its associated vacuum energy and the implications on the cosmic evolution.

{The interrelationship between the Higgs mechanism and gravity in its different formulations has been object of intensive research and has been considered in a variety of papers, e.g. recently in \cite{CQG33,PRD89} and previously: see e.g.  \cite{Dehnen92,PRD80,Brazphys40} and references therein. In the context of inflationary models the Higgs field can also play different rols\,\cite{PLB2008,Germani2010,MasinaPRL108,Tsujikawa2013,Leekoreanphys,Tahmasebzadeh2016}. Besides, the Higgs-induced spectroscopic shifts near strong gravity sources has been demonstrated in\,\cite{onfrioPRD82,PRD85}. Interesting connections between cosmological features of the Higgs boson and LHC physics have also been elucidated\, \cite{PRL110,PRD88}, and also in regard to astrophysical observations \cite{EPJC2015,APJ2014}. }

{Here, in our search for alternative frameworks that could explain the origin of the Higgs sector, we consider the theoretical possibility that the very structure of the Higgs potential is dictated by a real feedback between the particle physics world and the gravitational interactions}. Through this ``communicative ansatz'' we try to find a possible, more physical, explanation for the origin of the Higgs potential that is minimally satisfactory in the two large domains (Particle Physics and General Relativity/Cosmology) which have traditionally remained isolated from one another.  Specifically, in this work we explore an extend Jordan-Fierz-Brans-Dicke (``BD'' for short) type  of gravitational theory\,\cite{JF,BD}, in which apart from including the usual non-minimal scalar-tensor interaction\,\cite{ScalarTensor} for the BD-field, $\psi$, we introduce an interaction term between $\psi$ and a second scalar, $\phi$, which will play the role of Higgs boson after we determine self-consistently the form of its effective potential. {Remarkably, this is possible if $\phi$ interacts non-minimally with curvature and if its kinetic term interacts with the Ricci tensor, thereby through a derivative interaction with gravity\, \cite{Amendola93,Capozziello99}}.  We do not address here higher order gravitational theories, as in our case we limit ourselves to generalized forms of Einstein's gravity involving only the first power of the Ricci tensor and scalar, but we extend our study to the case of Grand Unified Theories (GUT's), where more than one type of Higgs field is involved.

The kind of cosmological solutions that we search for in order to fix self-consistently the Higgs potential are the simplest possible ones, namely the power-law solutions of the cosmological equations in a Friedmann-Lema\^{\i}tre-Robertson-Walker (FLRW) background. {These scaling solutions are interpreted as representing asymptotic states of the different phases of the cosmic evolution\, \cite{Amendola99,Guo2003,Granda,GrandaIJMPD22,MPLA28,DolgovJCAP10,JCAPO3,Kumar2012,Sharif82,phiCDM16}. The interpolation between the asymptotic states can only be described in terms of more general solutions, but in cosmology we are mostly interested in  phases characterized by a stationary equation of state (EoS), such as e.g. the de Sitter, the radiation-dominated and the matter-dominated epochs, rather than in the transitory regimes connecting the various stationary EoS phases.} We do not study the more general solutions connecting the asymptotic regimes since it is a much more complicated problem. However, the possibility to motivate the structure of the Higgs potential by picking out the cosmological solutions that are necessary for a correct account of the cosmic history is from our point of view already an interesting feature.

As we shall see, by requiring that the power-law solutions of this  model  are consistent with observations (i.e. not departing significantly from standard cosmology and a slowly time-varying Newtonian coupling), we find that the effective potential associated to $\phi$ may adopt the Higgs potential form under suitable conditions. Remarkably, the resulting vacuum expectation value is slowly evolving with the cosmic expansion, and such a mild time evolution could be responsible for the {possible time variation of the fundamental constants -- see \cite{Preface} for a short introduction and\,\cite{Uzan2011,Chiba2011,CalmetKeller,PlanckConstants2015} for more details. In particular, in \cite{PRD70,PRD77} the environmental dependence of masses and coupling constants through a quintessence field coupled to matter and gauge field is investigated. Let us also mention that a rather general class of dynamical vacuum models exist in the literature pointing out the possibility of variable masses and couplings -- see e.g.\,\cite{FritzschSola,FritzschSola2015,FritzschNunesSola} as well as other related approaches \cite{Flambaum2015} and references therein}\footnote{Some of these models have been recently confronted to the cosmological data in\,\cite{ApJL2015,DModels15,RVMs2016,DVMs2016,phiCDM16}. For a review, see \cite{JSPReview2016}.}.

 Another intriguing result of our work is that we find a possible link between the parameters of the Higgs potential, the gravitational coupling and the value of the CC, which is reminiscent of old insightful ideas on this subject by Bjorken \,\cite{Bjorken2001a,Bjorken2001b,Bjorken2010}, who made an interesting attempt to link the electroweak and gravitational scales.

The summary content of this paper is as follows. In Sect. II we briefly review Brans-Dicke gravity. In Sect. III we consider the extended BD formulation in the SM context and the ``determination'' of the Higgs potential. The implications on the time evolution of the fundamental constants and on the CC problem are addressed in Sect. IV, whereas in Sect. V we extend these ideas to Grand Unified Theories, taking $SU(5)$ as a prototype model. In the last section we deliver a closing discussion and our conclusions. In an appendix we collect some technical details for the analysis of the (more cumbersome) GUT case.

\section{Brans-Dicke gravity}

The Brans-Dicke (BD) theory\,\cite{JF,BD} is the first historical attempt to extended GR to accommodate variations in the Newtonian coupling $G$. A generalization of it has led to a wide panoply of scalar-tensor theories since long ago\,\cite{ScalarTensor}. The BD gravity is characterized by the scalar BD field $\psi$ coupled to the Ricci scalar of curvature, $R$, and by  a single  (dimensionless) constant parameter, $\omega$, in front of the kinetic term of $\psi$.
The original BD-action reads as follows\,\footnote{Here we use  metric and curvature conventions as in \cite{MTW73,LiddleLyth2000,AmendolaTsujikawa2010}.}:
\begin{eqnarray}
S_{\rm BD}=\int d^{4}x\sqrt{-g}\left[\frac{1}{16\pi}\left(R\psi-\frac{\omega}{\psi}g^{\mu\nu}\partial_{\nu}\psi\partial_{\mu}\psi\right)-\rL\right]+\int d^{4}x\sqrt{-g}\,{\cal L}_m(\chi_i,g_{\mu\nu})\,. \label{eq:BDaction}
\end{eqnarray}
The last term of this action stands for the matter action $S_{m}$, which is constructed from the Lagrangian density of the matter fields, $\chi_i$. Notice that there is no potential for the BD-field $\psi$ in the original BD-theory, and the dynamics of $\psi$ is such that $\psi(t_0)=1/G$ at present ($t=t_0$), where $G$ is the current Newtonian coupling. Therefore, $\psi$ has dimension-2 (i.e. mass dimension squared) in natural units. The effective value of $G$ is thus given by $1/\psi$.  The BD-action reduces to GR in the limit $\omega\to\infty$ (see below). Let us recall the corresponding field equations of motion after performing variation with respect to both the metric and the scalar field $\psi$. The result is the following:
\begin{equation}\label{eq:BDFieldEquation1}
\psi\,G_{\mu\nu}+\left(\Box\psi +\frac{\omega}{2\psi}\left(\nabla\psi\right)^2\right)\,g_{\mu\nu}-\nabla_{\mu}\nabla_{\nu}\psi-\frac{\omega}{\psi}\nabla_{\mu}\psi\nabla_{\nu}\psi=8\pi\left(\,T_{\mu\nu}-g_{\mu\nu}\rL\right)
\end{equation}
and
\begin{equation}\label{eq:BDFieldEquation2a}
\Box\psi-\frac{1}{2\psi}\left(\nabla\psi\right)^2+\frac{\psi}{2\omega}\,R=0\,,
\end{equation}
where we have assumed that both $\omega$ and $\rL$ are constants. To simplify the notation we have written  $(\nabla\psi)^2\equiv g^{\mu\nu}\nabla_{\mu}\psi\nabla_{\nu}\psi$.
In the first field equation, $G_{\mu\nu}=R_{\mu\nu}-(1/2)Rg_{\mu\nu}$ is the Einstein tensor, and  on its r.h.s. $T^{\mu \nu}=(2/\sqrt{-g})\delta S_{m}/\delta g_{\mu\nu}$ is the energy-momentum tensor from matter. We can use the trace of Eq.\,(\ref{eq:BDFieldEquation1}) to eliminate $R$ from (\ref{eq:BDFieldEquation2a}). Defining $T\equiv T^{\mu}_{\mu}$, equation (\ref{eq:BDFieldEquation2a}) can be brought to the simpler form
\begin{equation}\label{eq:BDFieldEquation2}
\Box\psi=\frac{8\pi}{2\omega+3}\,\left(T-4\rL\right)\,.
\end{equation}
We can immediately recognize from (\ref{eq:BDFieldEquation2}) that for $\omega\to\infty$ the solution is $\psi=$const. and,  as noted before, we then recover GR by selecting that constant to be the current value of $1/G$. With this choice, Eq.\,(\ref{eq:BDFieldEquation1}) boils down to the usual Einstein's equations and, of course, Eq.\,(\ref{eq:BDFieldEquation2a}) becomes trivial.

We have included a possible CC term or vacuum energy density, $\rL$, in the BD-action (\ref{eq:BDaction}). The quantum matter fields usually induce an additional, and very large, contribution to $\CC$. This is of course the origin of the CC problem\,\cite{CCProblem,JSPReview2013}. We shall come back to this issue later on after we extend the BD action in the next section.

The BD-action (\ref{eq:BDaction}) is given in the Jordan frame. After a conformal transformation of the metric, i.e. a metric transformation of the form $g_{\mu\nu}=\Omega(x)\bar{g}_{\mu\nu}$, in which $\Omega(x)\equiv e^{2\Theta(x)}$ is a local, positive, smooth function of the general coordinates, we can bring the BD-action into the Einstein frame. Indeed, upon choosing $\Omega(x)=\psi_0/\psi(x)$, for some fixed value $\psi_0$,  and defining a new scalar field $\bar{\psi}$ through $\bar{\psi}_0\,\partial\Theta/\partial\bar{\psi}=\sqrt{4\pi/(3\omega+2)}$ (which allows to put $\Theta$ as a function of $\bar{\psi}$), we find $\ln(\psi/\psi_0)=-\sqrt{16\pi/(3\omega+2)}\,\left(\bar{\psi}/\bar{\psi}_0\right)$, since $\omega=$const. (as in the original BD proposal). Finally, introducing a potential for $\bar{\psi}$ as $U(\bar{\psi})\equiv\Omega^2\rL$, the action (\ref{eq:BDaction}) becomes
\begin{eqnarray}
S_{\rm BD}=\int d^{4}x\sqrt{-\bar{g}}\left[\frac{1}{16\pi}\bar{R}-\frac12\left(\bar{\nabla}\bar{\psi}\right)^2- U(\bar{\psi})\right]+\int d^{4}x\sqrt{-\bar{g}}\,\Omega^2\,{\cal L}_m(\chi_i,\Omega\,\bar{g}_{\mu\nu})\,, \label{eq:BDactionEinstein}
\end{eqnarray}
where the over-bars denote quantities expressed in the barred metric, except $\bar{\psi}$ which denotes the aforementioned redefinition of $\psi$ that brings it in canonical form in the transformed action. As we can see, the new action after the conformal transformation became expressed in the Einstein frame since the scalar-tensor interaction present in the original action (Jordan frame) has disappeared and the normal Hilbert-Einstein term is obtained. However, the matter action now depends on the new metric and the conformal factor, $\Omega$, and therefore the matter fields $\chi_i$ do not follow the geodesics of the new matric. From this point of view the conformally transformed action is not physically equivalent to GR, which is of course a rephrasing of the fact that the initial BD-gravity action in the Jordan frame, Eq.\,(\ref{eq:BDaction}), is not equivalent to GR. The equivalence is only possible when $\psi=$const. ($\omega\to\infty$), as we have seen above.

\section{Extended Brans-Dicke action in the SM context }\label{set:BDinSM}
We now extend the Brans-Dicke type action (\ref{eq:BDaction}) with the BD-field, $\psi$, by including a new scalar field, $\phi$, with canonical dimension $1$ in natural units. The latter will eventually play the role of the Higgs boson field, as we shall see. We assume that there is a coupling term between these scalar fields, as well as non-minimal interactions with gravity, which we shall specify in a moment. We assume that the new BD-action includes all of the terms of the SM of strong and electroweak interactions, in particular the matter fields, but we are mainly interested on the part of this action comprising the dynamics and interaction between the scalar fields, as well as their couplings to gravity. The total action that we propose in this generalized BD-approach is:
\begin{eqnarray}
S=\int d^{4}x\sqrt{-g}\left[\frac12\,R\psi-\frac{\omega}{2\psi}g^{\mu\nu}\partial_{\nu}\psi\partial_{\mu}\psi+ \eta\phi^{2}\psi +\xi R\phi^{2}-\frac{1}{2} g^{\mu\nu}\partial_{\mu}\phi\,\partial_{\nu}\phi \right. \  \nonumber\\ +\left.\frac{1}{\phi^{2}}S_{\mu\nu}\partial^{\mu}\phi\partial^{\nu}\phi -V(\phi)\right]+S_m\,.  \label{eq:SMBDaction}
\end{eqnarray}
{Let us note that in this extension of the BD-theory we continue not ascribing any potential to the $\psi$ field. Therefore, we do not consider spontaneous symmetry breaking in the BD sector, as we wish to remain as close as possible to the original theory\,\cite{BD}.} The above action is invariant under global conformal (Weyl) transformations in what concerns the metric and scalar fields coupled to the gravitational part. However, after a local conformal transformation the action changes. In particular, following a similar procedure as in the previous section, it can be brought to the Einstein frame. In such frame it can be made arbitrarily close to the GR action because the field $\psi$ involved in the transformation changes very slowly with time. As we will see, this feature is tantamount to requiring a large value of the BD-parameter $\omega$. To fully retrieve GR, however, we need to switch off the remaining dimensionless parameters. We will assume that $\psi$ and $\phi$  do not change at all in space. In fact, to avoid conflict with the weak equivalence principle, the variability of $\psi$
in the BD-theory, and generalizations thereof, was originally restricted to a mild time evolution, $\psi=\psi(t)$, with no
spatial dependence\,\cite{Damour2012}. We shall address the possible time variation of $\psi$ and $\phi$ soon.

{As far as the extra terms in the above generalized BD-action, it includes the new tensor
\begin{equation}\label{eq:SmunuTensor}
S_{\mu\nu}\equiv\varsigma R_{\mu\nu}-\frac{\theta}{2}g_{\mu\nu}R\,,
\end{equation}
so as to make allowance for generalized derivative couplings of $\phi$ with gravity. The couplings involved in Eq.\,(\ref{eq:SmunuTensor}) were first studied in\,\cite{Amendola93,Capozziello99} and they turn out to be essential in this work since they eventually shape the structure of the Higgs boson self-coupling in the potential  that will be derived later on -- see Eqs.\,(\ref{eq:potential-phi})-(\ref{eq:Upsilon}). }

For the special case when the two parameters are equal, $\theta=\varsigma$, the tensor $S_{\mu\nu}$  becomes proportional to the Einstein tensor $G_{\mu\nu}$, but an important condition in our approach is that it should not coincide exactly with the Einstein tensor; namely, when $\varsigma=\theta$ these parameters cannot be equal to one, and in general $\varsigma$ and $\theta$ are different. At the same time, for convenience we have rescaled the BD-field of (\ref{eq:SMBDaction}) as compared to the original action (\ref{eq:BDaction}), namely $\psi\to 8\pi\psi$, such that now the effective gravitational coupling at any time of the cosmic expansion is furnished by
\begin{equation}\label{eq:Geff}
G_{\rm eff}(t)=\frac{1}{8\pi\psi(t)}.
\end{equation}
Equivalently, the effective value of the (reduced) Planck mass squared at any time is just given by the BD-field:
\begin{equation}\label{eq:MPeff}
M_P^2(t)=\psi(t)\,.
\end{equation}
At $t=t_0$ (current time) we have $M_P(t_0)\equiv M_P=1/\sqrt{8\pi G}\simeq2.43\times 10^{18}$ GeV.

Apart from the BD-parameter $\omega$ and the parameters defining the tensor (\ref{eq:SmunuTensor}), we have $\xi$ and $\eta$. All of them are dimensionless. Parameter $\xi$ stands for the non-minimal coupling of the field $\phi$ with the Ricci scalar, and $\eta$ is the coupling between the two scalar fields $\psi$ and $\phi$. However, let us emphasize that while $\psi$ couples to gravity in the standard, scalar-tensor form, $\phi$ couples to gravity non-minimally both non-derivatively  and derivatively\,\cite{Amendola93,Capozziello99}. The above action reduces to GR in the limit $\omega\to\infty$ when the remaining couplings go to zero. In general, however, all the mentioned couplings take on non-vanishing values, and as we shall see there are many possibilities compatible with the observed phenomenology.

Finally,  note that we leave the scalar potential $V(\phi)$ for the new scalar field $\phi$ unspecified at this point. In fact, our intention is to show that its form will be that of the Higgs potential after we impose the appropriate conditions. Needless to say any constant contribution, such as $\rL$ in Eq. (\ref{eq:BDaction}), can  be absorbed as part of  $V(\phi)$.

\subsection{Equations of motion}\label{sect:eqmotionSM}

Upon taking the variational derivatives of the action (\ref{eq:SMBDaction}) with respect to the metric and the two scalar fields $\phi$ and $\psi$ we obtain the field equations, which generalize the ones provided in the previous section and include one additional equation for the new filed $\phi$. Performing first the variation with respect to the metric, we find:

\begin{eqnarray}
\left(G_{\mu\nu}+D_{\mu\nu}\right)\psi+\frac{\omega}{2\psi}g_{\mu\nu}(\nabla\psi)^2-\frac{\omega}{\psi}\nabla_{\nu}\psi\nabla_{\mu}\psi -g_{\mu\nu}\eta\phi^{2}\psi-\nabla_{\nu}\phi\nabla_{\mu}\phi \nonumber\\ +\frac{1}{2}g_{\mu\nu}\left(\nabla_{\alpha}\phi\nabla^{\alpha}\phi\right)+2\xi\left(G_{\mu\nu}+D_{\mu\nu}\right)\phi^{2} +2\varsigma \left\lbrace \frac{1}{2}g_{\mu\nu}\nabla_{\beta}\nabla_{\alpha}\left(\phi^{-2}\nabla^{\alpha}\phi\nabla^{\beta}\phi\right)\right.\nonumber\\ +\left.R_{\mu\alpha}\left(\phi^{-2}\nabla^{\alpha}\phi\nabla_{\nu}\phi\right)+R_{\nu\alpha}\left(\phi^{-2}\nabla^{\alpha}\phi\nabla_{\mu}\phi\right)-\frac{1}{2}g_{\mu\nu}R_{\alpha\beta}\left( \phi^{-2}\nabla^{\alpha}\phi\nabla^{\beta}\phi\right)\right. \  \nonumber\\ -\left.\frac{1}{2}\left(\nabla_{\mu}\nabla_{\beta}\left( \phi^{-2}\nabla_{\nu}\phi\nabla^{\beta}\phi\right)+\nabla_{\nu}\nabla_{\beta}\left( \phi^{-2}\nabla_{\mu}\phi\nabla^{\beta}\phi\right)\right)+\frac{1}{2}\square\left(\phi^{-2}\nabla_{\mu}\phi\nabla_{\nu}\phi\right) \right\rbrace \nonumber\\ -\theta\left\lbrace \left(G_{\mu\nu}+D_{\mu\nu}\right)\left(\phi^{-2}\nabla_{\alpha}\phi\nabla^{\alpha}\phi\right)+R \left(\phi^{-2}\nabla_{\mu}\phi\nabla_{\nu}\phi\right)\right\rbrace= T_{\mu\nu}^{m}-g_{\mu\nu}V(\phi)  , \label{VariationMetric}
\end{eqnarray}
where we have defined $D_{\mu\nu}\equiv g_{\mu\nu}\square-\nabla_{\mu}\nabla_{\nu}$. The corresponding variations with respect to $ \phi$ and $\psi$ lead respectively to the additional field equations
\begin{eqnarray}
\Box\phi+\frac{2}{\phi^{3}}S_{\mu\nu}\nabla^{\mu}\phi\nabla^{\nu}\phi-\frac{2}{\phi^{2}}\left\lbrace\left(\nabla^{\mu} S_{\mu\nu}\right)\nabla^{\nu}\phi+S_{\mu\nu}\left(\nabla^{\mu} \nabla^{\nu}\phi\right)\right\rbrace+2\left(\eta\psi+\xi R\right)\phi-\frac{dV}{d\phi}=0\,,\nonumber\\ \label{Variationphi}
\end{eqnarray}
and
\begin{equation}
\Box\psi-\frac{1}{2\psi}\left(\nabla\psi\right)^2+\frac{\psi}{2\omega}\,R+ \frac{\eta}{\omega}\psi\, \phi^{2}=0
\label{Variationpsi}
\end{equation}
We can check that in the vanishing limit of all the generalized couplings  $\xi=\eta=\theta=\varsigma=0$ (except $\omega$) the first equation above, Eq.\,(\ref{VariationMetric}), boils down to (\ref{eq:BDFieldEquation1}) -- recall, however, our rescaling $\psi\to 8\pi\psi$. Similarly, equation (\ref{Variationpsi}) reduces to (\ref{eq:BDFieldEquation2a}). Finally, Eq.\,(\ref{Variationphi}) becomes a decoupled equation for $\phi$ in the same limit, as expected.

Now, in order to generate specific equations of motion in the cosmological context, we adopt the FLRW background metric:
\begin{equation}
ds^{2}=-dt^{2}+a^{2}(t)\left(\frac{dr^{2}}{1-kr^{2}}+r^{2}d\Omega^{2}\right)\,. \label{eq:frwmetric}
\end{equation}
In practice we shall focus only on the spatially-flat case, i.e. $k=0$.

The above field equations (\ref{VariationMetric})-(\ref{Variationpsi}) can now be written out in the flat FLRW metric (\ref{eq:frwmetric}), in which $a(t)$ is the scale factor. From it we can define the expansion rate or Hubble function: $H=\dot{a}/a$. Let us neglect hereafter the matter contribution since  we wish to focus on the regime where the scalar fields dominate. In this regime the explicit form of the field equations can be worked out with the following results:
\begin{eqnarray}
{{3}H^{2}\psi}+{{3}H{\dot{\psi}}}-\frac{\omega}{2}\frac{\dot{\psi}^{2}}{\psi}+\eta\phi^{2}\psi-\frac{1}{2}\dot{\phi}^{2}-V(\phi)+6\xi H^{2}\phi^{2}+12\xi H \dot{\phi}\phi-{9}\theta H^{2}\frac{\dot{\phi}^{2}}{\phi^{2}}- \nonumber \\
 6\left(\theta-\varsigma\right)\dot{H}\frac{\dot{\phi}^{2}}{\phi^{2}}+6\left(\theta-\varsigma\right)H\frac{\dot{\phi}\ddot{\phi}}{\phi^{2}}-6\left(\theta-\varsigma\right){H}\frac{\dot{\phi}^{3}}{\phi^{3}}=0\,, \ \ \ \ \label{eq:EoM-metric}
\end{eqnarray}
\begin{eqnarray}
\ddot{\phi}+3 H \dot{\phi}-12\xi\dot{H}\phi - 24\xi H^{2}\phi+\frac{d V}{d\phi}+6 \left(2\theta -\varsigma\right) H^{2}\left(\frac{\ddot{\phi}}{\phi^{2}}-\frac{\dot{\phi}^{2}}{\phi^{3}}\right)+18 \left( 2 \theta -\varsigma \right) {H}^{3}\frac{\dot{\phi}}{\phi^{2}}+ \nonumber\\
6 \left(7 \theta -5 \varsigma \right) H \dot{H}\frac{\dot{\phi}}{\phi^{2}}+6 \left(\theta - \varsigma \right)\ddot{H}\frac{\dot{\phi}}{\phi^{2}}+6 \left(\theta -\varsigma\right)\dot{H}\left(\frac{\ddot{\phi}}{\phi^{2}}-\frac{\dot{\phi}^{2}}{\phi^{3}}\right)-{2}\eta \psi\phi= 0\,,\phantom{XXX}\label{eq:EoM-phi}
\end{eqnarray}
and
\begin{equation}
3\dot{H}+6{H}^{2} - \omega \frac{\ddot{\psi}}{\psi}+\frac{ \omega}{2} \frac{\dot{\psi}^{2}}{{\psi}^{2}}-3H\omega\frac{\dot{\psi}}{\psi}+ \eta {\phi} ^{2}=0\,.  \ \ \ \ \ \ \ \ \ \ \label{eq:EoM-psi}
\end{equation}
It is worthwhile to mention here that by taking into account the above equations of motion, i.e. Eqs. (\ref{eq:EoM-metric}), (\ref{eq:EoM-phi}) and (\ref{eq:EoM-psi}), one can check  that the local covariant conservation law of the energy-momentum tensor ($\nabla^{\mu} T_{\mu \nu}=0$) is satisfied, where $T_{\mu \nu}=(-2/\sqrt{-g})\delta S/\delta g^{\mu\nu}$ is computed from the action of Eq.\,(\ref{eq:SMBDaction}).

The general solution of the above coupled system of (non-linear) differential equations is of course rather complicated. However, in the relevant regime of scalar field dominance we may seek power-law solutions
\begin{equation}
H=\frac{p}{t} \ \ (p>0)\,, \ \ \ \ \ \phi(t)=\phi_{1}\left(\frac{t}{t_{1}}\right)^{\alpha}, \ \ \ \ \ \ \ \ \psi(t)=\psi_{1}\left(\frac{t}{t_{1}}\right)^{\gamma}\,.\label{eq:power law}
\end{equation}
Such scaling ansatz can be useful to search for asymptotic behaviors of the equations in the various phases of the cosmic evolution. Here $p,\alpha,\gamma$ are dimensionless parameters, and $\phi_{1}$ and $\psi_{1}$ are the  values of the scalar fields at some reference time $t=t_1$ in the past when the power law solutions start to hold good for such period. In particular $\gamma$ controls the ``running'' of the Planck mass (\ref{eq:MPeff}) -- and hence of Newton's gravitational coupling, Eq.\,(\ref{eq:Geff})-- in the context of BD-gravity. Obviously we expect $|\gamma|$ to be small. Different periods of the cosmic evolution correspond to different scaling solutions. For instance,  $p=1/2$ and $p=2/3$ would represent cosmological states in the radiation-dominated and matter-dominated epochs, respectively. Usually each period is characterized by the dominance of a particular energy density $\rho$ (radiation, matter, vacuum energy or some form of dark energy), which is related to the Hubble function through Friedmann's equation $H^2=(8\pi G/3)\rho$, in which the dominant density $\rho$ satisfies the local conservation law $\dot{\rho}+3H(\rho+p)=0$. Defining the equation of state (EoS) parameter $w=p/\rho$, it follows that each scaling period (\ref{eq:power law}) is characterized by the EoS value
\begin{equation}\label{eq:EoS}
w=-1-\frac{\dot{\rho}}{3H\rho}=-1-\frac23\,\frac{\dot{H}}{H^2}=-1+\frac{2}{3p}\,.
\end{equation}
As expected, this yields $w=1/3$ and $w=0$ for radiation (relativistic matter) and non-relativistic matter, respectively. Notice also that for $p>1$ we find $w<-1/3$, which corresponds to a period of acceleration and therefore it can be associated to an inflationary phase. In particular, for $p\to\infty$ we recover the EoS of the vacuum energy $w=-1$.  We may easily confirm from the first scaling relation in (\ref{eq:power law}) that fort $p>1$ we are in a phase of positive acceleration:
\begin{equation}\label{eq:inflation}
\frac{\ddot{a}}{a}=\dot{H}+H^2=\frac{p(p-1)}{t^2}>0\ \ \ \ (\text{for}\ p>1)\,.
\end{equation}

Using the above scaling ansatz we will now show that it is possible to expect that the Universe during such stationary EoS periods ($p=$const, $w=$const.) is characterized by an effective potential for $\phi$ which takes on the Higgs potential form under appropriate conditions. Let us derive the form of such potential in a self-consistent form. Substituting equations (\ref{eq:power law}) in the field equation (\ref{eq:EoM-metric}) provides a first explicit expression for the potential as a function of the cosmic time:
\begin{eqnarray}
V=\frac{\psi_{1}}{ t_{1}^{\gamma}}\left({3p(p+\gamma)}- \frac{\omega}{2} \gamma^{2}\right)t^{\gamma-{2}}-9 p^{2} \alpha ^{2}  \theta t^{-4}+\frac{\phi_{1}^{2}}{t_{1}^{2\alpha}}\left(6\xi p\left(p+2\alpha\right)-\frac{1}{2}\alpha^{2}\right) t^{2\alpha-2}\nonumber \\ +\ \eta \frac{\phi_{1}^{2} \psi_{1}}{t_{1}^{{2}\alpha+\gamma}}t^{{2}\alpha+\gamma}. \label{eq:EoM-metric-1}
\end{eqnarray}
Inserting also equations (\ref{eq:power law}) into the second field equation (\ref{eq:EoM-phi}) we find an expression for the derivative of the potential, $dV/d\phi=f(t)$, in which the function $f(t)$ is known. This expression can be integrated as $V=\int f(t)\dot{\phi}dt=\alpha\,(\phi_1/t_1)\int f(t)\,(t/t_1)^{\alpha-1}\,dt\ $ after using once more (\ref{eq:power law}). For simplicity we set the integration constant to zero (we shall discuss the CC problem later on). In this way we are led to an alternative form of the potential as a function of time:
\begin{eqnarray}
V=-\frac{ \alpha \phi_{1}^{2}}{\left({2}\alpha-2\right)}\left(\alpha\left(3p+\alpha-1\right)-12\xi p\left(2p-1\right)\right)\frac{t^{{2}\alpha-2}}{ t_{1}^{{2}\alpha}}+ \frac{{2}\eta \psi_{1} \phi_{1}^{2} \alpha}{\left({2}\alpha+\gamma\right)}\frac{t^{{2}\alpha+\gamma}} {t_{1}^{{2}\alpha+\gamma}}\nonumber \\  -\  \frac{9}{2} p\alpha^{2}\sigma t^{-4} , \label{eq:EoM-phi-1}
\end{eqnarray}
in which $\sigma=\left(3\theta -2\varsigma\right)p -\left(2\theta-\varsigma\right)p^{2} - \left(\theta-\varsigma\right)$. In the particular case $\theta=\varsigma$ (characterized by $S_{\mu\nu}$ being proportional to $G_{\mu\nu}$), we have $\sigma=\varsigma p\left(1-p\right)$. By repeating the same procedure with the third field equation (\ref{eq:EoM-psi}), which does not depend on the potential, we meet now a consistency relation:
\begin{equation}
\eta \phi_{1}^{2} t^{{2}\alpha}= \left[\omega \gamma \left(\frac{\gamma}{2}+3p-1\right)+{3 p \left(1-{2}p\right)}\right] t_{1}^{{2}\alpha} t^{-2}.  \label{eq:EoM-psi-1}
\end{equation}
This equation emerges directly from considering BD-gravity, as it is generated from the variation of the action with respect to the BD-field $\psi$. Thanks to the nonvanishing coupling $\eta$ between the two scalar fields, such relation becomes essential to pinpoint a unique value for the parameter $\alpha$. The only choice that fulfills Eq. (\ref{eq:EoM-psi-1}) is $\alpha=-1$. We are now ready to determine a more precise form of the potential.


\subsection{Determination of the Higgs potential}\label{sect:HiggsPotentialSM}

Inserting  $\alpha=-1$ in equations (\ref{eq:EoM-metric-1}) and (\ref{eq:EoM-phi-1}), we find:
\small\begin{eqnarray}
V=\left(\phi_{1}^{2} t_{1}^{2}\left[6\xi p\left(p-2\right)-\frac{1}{2}\right]-9 p^{2}\theta\right) t^{-4}+\frac{\psi_{1}}{ t_{1}^{\gamma}} \left({3p(\gamma+p)}-\frac{\omega}{2} \gamma^{2}+\eta \phi_{1}^{2}t_{1}^{2}\right) t^{\gamma-{2}}\label{eq:EoM-metric-2}
\end{eqnarray}
and
\begin{eqnarray}
V=-\frac{\psi_{1}}{t_{1}^{\gamma}}\left(\frac{{2}\eta \phi_{1}^{2}t_{1}^{2} }{ \gamma-2}\right)
 t^{\gamma-{2}}-\frac{1}{2}\left[\frac{\phi_{1}^{2}t_{1}^{2}}{2}\left(12\xi p\left(1-2p\right)-3p+2\right)+9 \sigma p\right] t^{-4}\,, \label{eq:EoM-phi-2}
\end{eqnarray}
respectively.
The above two forms of the potential for $\phi$ can be made mathematically consistent only in two  ways, although only one option will be physically acceptable. Suppose that  $t^{\gamma-{2}}$ and $t^{-4}$ were to be the same power, which would entail $\gamma=-2$ . This option is, however, ruled out on physical grounds since the third scaling relation in Eq.\,(\ref{eq:power law}) would imply $\psi(t)\varpropto t^{-2}$, and then from Eq.\,(\ref{eq:Geff}) we would have $G(t)\varpropto t^{2}$. Obviously, such a rapid evolution of the gravitational coupling constant cannot be accepted. Among other things it would completely spoil the primordial BBN stages of the cosmic evolution and subsequently those of structure formation at late time. By excluding this option, we are only left with  the second solution, which is to treat $t^{\gamma-{2}}$ and $t^{-4}$ as two independent powers. In this case, the corresponding coefficients of  (\ref{eq:EoM-metric-2}) and (\ref{eq:EoM-phi-2}) should match. This leads to the following consistency conditions:
\begin{equation}
\phi_{1}^{2}t_{1}^{2}=\frac{6\sigma-12p\theta}{12\xi+1} \label{eq:restriction-sigma}
\end{equation}
and
\begin{equation}
- \frac{2\gamma \eta}{\gamma-2} \phi_{1}^{2} t_{1}^{2} = 6p\left(\gamma + p\right)- \omega \gamma^{2}.   \label{eq:restriction-eta}
\end{equation}
The last two equations clearly correlate the signs and sizes of the various parameters. However, some of the parameters are restrained by important physical conditions.

For example, we expect that $\gamma$ should have a rather small absolute value because $\psi\propto1/G_{\rm eff}\propto t^{\gamma}$ and in this way the effective time-evolving value of $G_{\rm eff}$, Eq.\,(\ref{eq:Geff}), can be consistent with observations. Observationally we know that $|\gamma|\lesssim10^{-3}$\,\cite{Uzan2011,Chiba2011} but there is no surety about its sign, see e.g. \cite{Li13}. On the other hand, $\omega$ must be large and in fact it can be as large as desired since the limit $\omega\to\infty$ is, as we know, a necessary condition to retrieve GR. Observationally we know that it is so. In \cite{Avilez}, for instance, the BD parameter $\omega$ is constrained to satisfy $\omega >890$ at the $99\%$ confidence level within the canonical BD formulation and on the basis of cosmological data (essentially on CMB)\,\footnote{At the much lower scale of the Solar System, however, the Cassini mission provided accurate enough data on the parameters of the PPN formalism to infer $\omega>40000$ at $2\sigma$ c.l.\,\cite{Cassini}.}.

We are now ready to derive the final expression of the effective potential for $\phi$.  Considering Eqs. (\ref{eq:EoM-metric-2}) and (\ref{eq:EoM-phi-2}), under the consistency conditions (\ref{eq:restriction-sigma}) and (\ref{eq:restriction-eta}) together with the scaling laws (\ref{eq:power law}), we obtain the following form for the effective potential:
\begin{equation}
V=-\frac{{2}\eta\,\psi_{1}}{\gamma-{2}}\left(\frac{t}{t_1}\right)^{\gamma}\,\phi^{2}+\frac{ \Upsilon}{\phi_{1}^{4}t_{1}^{4}}\phi^{4}\simeq -\frac{{2}\eta \psi_{1}}{\gamma-{2}} \phi^{{2}}+\frac{\Upsilon }{\phi_{1}^{4}t_{1}^{4}}\phi^{4}\,, \label{eq:potential-phi}
\end{equation}
where in the approximate relation on the second equality we have taken into account that $|\gamma|$ is very small, and thus the coefficient of $\phi^2$ remains essentially constant. We will reconsider the mild time evolution effects later on, but at this point it suffices to note that the obtained potential is essentially a Higgs-like potential for $\phi$, with the coefficient $\Upsilon$  in the quartic coupling given by
\begin{eqnarray}
\Upsilon=\frac{3}{12\xi +1}\left\lbrace 12 p\xi \left[p^{3}\left(\varsigma-2\theta\right)+p^{2}\left(5\theta-4\varsigma\right)+p\left(5\varsigma-6\theta\right)+2(\theta-\varsigma)\right]-\right. \ \ \ \ \ \ \ \ \ \ \ \nonumber\\ \left. p^{2}(\theta+\varsigma)+ p \left(2\varsigma-\theta\right)+\left(\theta-\varsigma\right)\right\rbrace\  . \label{eq:Upsilon}
\end{eqnarray}
{As we can see from this expression, the structure of the Higgs self-coupling in the present framework is indeed shaped with the help of the coefficients involved in the nonminimal derivative coupling in the action (\ref{eq:SMBDaction}), see the $S_{\mu\nu}$ tensor in  Eq.\,(\ref{eq:SmunuTensor}). }

It is apparent that Eq.\, (\ref{eq:potential-phi}) can be interpreted as a Higgs type potential of generic form
\begin{equation}\label{eq:VHiggs}
V=\frac{\mu^{2}}{2}\phi^{2}+\frac{\lambda}{4}\phi^{4}  , \ \ \ \ \ \ \ \ \mu^2<0 \ \ \ \ \lambda>0\,,
\end{equation}
for appropriate values of the parameters $\mu^2$ (with dimension of mass squared) and $\lambda$ (the dimensionless quartic coupling). These values are determined here by the coefficients of the extended BD-action (\ref{eq:SMBDaction}) and by the form of the scaling solution (\ref{eq:power law}). This is tantamount to saying that the Higgs  form of the potential  is consistent, in our BD-framework, with the solutions that describe the basic epochs of the cosmic expansion. We shall discuss in a moment the connection of the BD-parameter $\omega$ with the parameters $p$ and $\gamma$ of the scaling solution.

The VEV of $\phi$, i.e. the location of the minimum of $V$, is given by $v=\sqrt{-\mu^2/\lambda}$ and hence the mass of the physical Higgs particle is simply given by
\begin{equation}\label{eq:HiggsMass}
M_H^2= \left.\frac{\partial^2 V}{\partial\phi^2}\right|_{\phi=v}=\mu^2+3\lambda\,v^2=-2\mu^2>0\,.
\end{equation}
Obviously $\mu^2<0$, and this is a key feature in order to determine the sign of some coupling constants in our BD-framework.  Thus, if we consider the coefficient of $\phi^{2}$ in the Higgs potential (\ref{eq:potential-phi}) it is clear that in order to trigger spontaneous symmetry breaking (SSB) of the EW symmetry  we must require it to be negative, and since $|\gamma|\ll1$ it follows that $\eta<0$.   Also from the fact that $\mu^2$ must be roughly identified with $\eta\psi_1$ in (\ref{eq:potential-phi}), and taking into account that $\psi_1\sim M_P^2$, we learn that $|\eta|$ must be a very small parameter in order to make contact with the EW theory (see next section for a detailed discussion).

Parameters $\gamma$ and $\omega$ are actually not independent at fixed $p$. Combining the consistency conditions (\ref{eq:EoM-psi-1}) and (\ref{eq:restriction-eta}) (with $\alpha=-1$, of course) we find after some algebra the simple relation:
\begin{equation}
\gamma^2\left(1+\omega\right)-\gamma\left(1+p\right)-2p=0\,, \label{eq:restriction-3}
\end{equation}
whose exact solution is
\begin{equation}
\gamma=\frac{1+p}{2\omega}\pm \sqrt{\frac{2p}{\omega}}\,\sqrt{1+\frac{(1+p)^2}{8p\omega}}\,. \label{eq:restriction-gammaEXACT}
\end{equation}
It is easy to see, however, that for $\omega\gg p\simeq 1/2$ and in general for ${\omega\gg\cal O}(1)$ (which is the natural situation, as we should not depart too much from GR) the approximate solution reads, in very good approximation:
\begin{equation}
\gamma\simeq -\sqrt{\frac{2p}{\omega}}  \label{eq:restriction-gamma}\,,
\end{equation}
where we have stood out the negative sign (despite both signs would be possible, in principle) for reasons to be discussed in a moment.
This equation is remarkable since it links in a specific way the expected large value of $\omega$ in BD-gravity with a possible mild ``running'' of the gravitational ``constant'', $G$, which is controlled by the coefficient $\gamma$ in the third scaling relation (\ref{eq:power law}). In particular, we can see from (\ref{eq:restriction-gamma}) that $\omega\to\infty$, if and only if $\gamma\to0$, which is the limit of GR in which $G=$const. We will come back to this point in the next section.

Furthermore, due to the relation (\ref{eq:restriction-gamma}), $\gamma$ depends on the specific epoch of expanding universe. For example, for the radiation epoch $p=1/2$, and for the matter-dominated epoch $p=2/3$. Inserting $\gamma^2\omega=2p$ --- from Eq. (\ref{eq:restriction-gamma}) -- on the $r.h.s$ of Eq. (\ref{eq:restriction-eta}) and baring in mind that $|\gamma|\ll1$ it amounts to the condition $\eta\gamma>0$ for $p>1/3$ (which involves all the aforementioned relevant epochs of the cosmic evolution). Combining this outcome with the previously noted $\eta<0$, it follows that  $\gamma$ must be a negative quantity and hence the reason for the minus sign in Eq. (\ref{eq:restriction-gamma}). Thus, the signs $\eta<0$ and $\gamma<0$ become uniquely fixed. We will retake the discussion of the important implications of these results later on.

Recall that we expect that $p$ should take at this stage a value near $p\simeq 1/2$ since the radiation epoch is the time when the EW phase transition must have occurred.  Later on the  Universe will be in transit to the matter-dominated epoch, characterized by $p=2/3$. However,  as we have mentioned in the introduction, we do not attempt to explain the transition from one scaling solution to the other, as this would imply to determine a more general (and certainly more complicated) class of solutions interpolating between them. Therefore, we limit ourselves at the moment to dwell on the regime close to the radiation epoch, in which by virtue of  Eq. (\ref{eq:restriction-gamma}) the condition  $\omega\gamma^2\simeq 1$ holds good.

There are of course a number of possibilities available for the structure of the potential, but for illustration we may assume $\varsigma=\theta$. In this case the coefficient (\ref{eq:Upsilon}) reduces to
\begin{equation}
\Upsilon=\frac{-3\varsigma p}{12\xi+1}\left[12p\xi \left(p^{2}-p+1\right)+2p-1\right].  \label{eq:Upsilon-simple}
\end{equation}
As remarked, we will be mostly interested in the radiation-dominated epoch, so we assume here $p=1/2$. Equation (\ref{eq:Upsilon-simple}) tells us that unless $\xi$ lies in the narrow negative range $-1/12<\xi<0$, the coefficient $\Upsilon$ above -- and for that matter the full quartic coupling in the potential (\ref{eq:potential-phi})-- remains positive provided  $\varsigma<0$.  Only if $\xi$ lies in that small negative interval the sign $\varsigma>0$ would be allowed, in principle. But in fact it is not so, as we next show. On combining equations (\ref{eq:restriction-sigma}) and (\ref{eq:restriction-eta}) we find the following relation: $\phi_{1}^{2}t_{1}^{2}=-6\varsigma p(p+1)/{(12\xi+1)}\simeq2p(3p-1)/(\gamma\eta)$, where we have also used (\ref{eq:restriction-gamma}). Since $\eta,\gamma<0$, and $p>1/3$, it turns out that the previous relation remains consistent, together with the positivity of the quartic coupling of the potential, only if $\xi>0$ and $\varsigma<0$. So the signs of all the coefficients are uniquely determined. Notice that this situation could not have been achieved if we had defined $\varsigma=\theta=\pm 1$ when we introduced the $S_{\mu\nu}$ tensor (\ref{eq:SmunuTensor}) with derivative couplings to $\phi$, which is one of the central pieces of our generalized BD-action (\ref{eq:SMBDaction}). While the case $\theta=\varsigma$ is, of course, not the most general, it already captures the basic features of our analysis.

\section{Vacuum energy and dynamics of fundamental constants}\label{sect:FundamentalConstants}

The problem of understanding the nature of the dark energy (DE) in the Universe\,\cite{AmendolaTsujikawa2010} is one of the most relevant ones in fundamental physics.  The original form of this problem is actually the famous cosmological constant problem\,\cite{CCProblem,JSPReview2013,JSPReview2016}. As indicated in the introduction, the electroweak vacuum energy is part of such theoretical conundrum, for the existence of the Higgs boson itself (as an experimentally accepted fact\,\cite{FindingH2012}) goes undissociated with the existence of the EW vacuum energy. So, from this point of view, the EW vacuum energy became recently an empirical ingredient of the CC problem, what makes such problem less formal and more real. In any whatever formulation, the DE or CC problem presents insurmountable theoretical difficulties and no simple solution can be expected shortly. We are certainly not addressing here an attempt at solving it. Our aim is much more modest. In what follows, however, we will explore an interesting possibility which might help to alleviate some of the encountered difficulties, and the approach can naturally be related with the class of generalized BD-gravity models that we are studying here. In addition, we will see that this framework may also impinge on the issue of the time variation of fundamental constants\,\cite{Uzan2011,Chiba2011,CalmetKeller,PlanckConstants2015}.

\subsection{Bjorken's approach to vacuum energy and the cosmological constant problem}

A decade and a half ago, J. Bjorken made an interesting proposal in which he tried to relate the parameters of the SM of particle physics (such as the gauge, Yukawa and Higgs self-couplings) with the two basic ones of the gravity world (namely Newton's gravitational coupling, or equivalently the Planck mass, and the cosmological constant). Details are given in \,\cite{Bjorken2001a,Bjorken2001b,Bjorken2010}. The idea is that the inner mechanisms capable of solving the hierarchy problem, i.e. why the quadratically divergent Higgs-boson mass remains at the EW scale, should take care of the cosmological constant problem as well. In such context the major parameters of the standard model are connected to the usual Planck scale  $m_P=1/\sqrt{G}\simeq 1.22\times 10^{19}$ GeV and to the millielectronvolt scale associated to the cosmological constant, $m_{\CC}\equiv\rL^{1/4}\sim 10^{-3}$ eV , through the large logarithm
\begin{equation}\label{eq:largelog}
 k\equiv\ln\frac{m_P^2}{m_{\CC}^2}\sim 145\,.
 \end{equation}
Bjorken suggested, in the aforementioned papers, that the reason why the gauge coupling $g$, the top quark coupling $h$ and the Higgs quartic coupling $\lambda$ are all perturbative, and hence small enough, is because they are related to the large log (\ref{eq:largelog}) through
\begin{equation}\label{eq:gaugelargelog}
 \frac{1}{g^2}\sim \frac{1}{h^2}\sim \frac{1}{\lambda}\sim \frac{1}{4\pi^2}\,\ln\frac{m_P^2}{m_{\CC}^2}=\frac{k}{4\pi^2}\,,
 \end{equation}
so that  $g^2/4\pi\sim \pi/k$ and the electromagnetic fine structure constant $\alpha_{\rm em}=g^2\sin^2{\theta_W}/4\pi$ are naturally small.
The next intriguing observation by Bjorken is that the VEV of the EW theory, $v\simeq 246$ GeV, is given roughly by the geometric mean of the Planck scale and the CC scale, that is to say $v^2\sim m_P\,m_{\CC}$. This ansatz is true but not quite, for we expect again (as also indicated by Bjorken) that it should hold good only up to dimensionless factors which may contain the above large log. We can see that it is more accurate to refine it as follows:
\begin{equation}\label{eq:kgeometricmean}
 v^2\simeq m_P\,m_{\CC}/k\simeq 6\times 10^4\, GeV^2\,,
 \end{equation}
 where $k$ reflects the large log (\ref{eq:largelog}). Needless to say, we do not aim so much at numerical precision here as to understand the orders of magnitude.  Such correction clearly insures a better agreement. Combined with (\ref{eq:gaugelargelog}), the previous formula can be reexpressed as
 \begin{equation}\label{eq:v2b}
 v^2\simeq\frac{g^2}{4\pi^2} m_P\,m_{\CC}\,.
 \end{equation}
In this form, it suggests that the more correct form of the original ansatz might be connected with a higher order correction. Any of these expressions is ultimately telling us an important message, which is at the heart of Bjorken's proposal, namely that $m_{\CC}=0 \Leftrightarrow v=0$, that is to say, the EW vacuum energy density of the SM is zero, if and only if, there is no cosmological constant.  Bjorken's ideas seem to point to something deep, for if a relation of that sort could be substantiated it would give a direct connection of the CC value with the electroweak Higgs condensate or VEV, what could perhaps shed some light on the issue discussed in the introduction on how to make compatible the EW sector (and the Higgs boson finding) with the measured value of the CC in cosmology.

We shall not further pursue this approach here, except to note that if Bjorken's ansatz (\ref{eq:kgeometricmean}) is admitted we can rewrite it as
 \begin{equation}\label{eq:v2lambda1}
 v^2=r_{Bj}\,m_P^2\,.
 \end{equation}
This relation defines a dimensionless and positive constant, $r_{Bj}>0$, which we may call  Bjorken's ratio, defining the squared quotient of the EW scale and the Planck scale. As it turns, it can be expressed in terms of only gravitational parameters, $m_P$ and the scale associated to the cosmological constant, up to a weighting factor:
\begin{equation}\label{eq:v2lambda}
 r_{Bj}\equiv{\frac{m_{\CC}}{k\,m_P}}\sim 10^{-33}\,.
 \end{equation}
Obviously, Bjorken's ratio is a very small number, but a natural one in this framework. As we shall see in the next section, it bears close relation with our parameter $\eta$ in the action (\ref{eq:SMBDaction}).

\subsection{Time variation of fundamental constants}

Let us further inquire on the structure of the effective potential (\ref{eq:potential-phi}), in which we want to retain now the original time-dependence:
\begin{equation}
V=-\frac{{2}\eta\,\psi_{1}}{\gamma-{2}}\left(\frac{t}{t_1}\right)^{\gamma}\,\phi^{2}+\frac{ \Upsilon}{\phi_{1}^{4}t_{1}^{4}}\phi^{4}\simeq {\eta\,\psi_{1}}\left(\frac{t}{t_1}\right)^{\gamma}\phi^{2}+\frac{ \Upsilon}{\phi_{1}^{4}t_{1}^{4}}\phi^{4}\,, \label{eq:potential-psi}
\end{equation}
where we recall that $|\gamma|\ll1$.
On comparing the previous expression of the potential with (\ref{eq:VHiggs}) and (\ref{eq:HiggsMass}) we see that the mass squared of the Higgs field is given by
\begin{equation}
M^2_{H}(t)=-4{\eta}\,\psi_{1}\left(\frac{t}{t_1}\right)^{\gamma}\simeq -4{\eta}\,\psi_{1}\left(1+\gamma\,\ln\frac{t}{t_1}\right)\,. \label{eq:Higgs mass}
\end{equation}
Being $\psi_1>0$, it obviously follows that $\eta<0$ (as it was noted) so as to warrant $M_H^2>0$.  In absolute value, the parameter $\eta$ must be very small. Specifically, if we neglect for this estimate the log term proportional to $\gamma$, we find
\begin{equation}\label{eq:etaestimate}
|\eta|\simeq \frac{M_H^2}{4\,\psi_1}\simeq 6.6\times 10^{-34}
\end{equation}
where $M_H\simeq 125$ GeV (from the observed value of the Higgs mass\,\cite{FindingH2012}) and $\psi_1\sim M_P^2$, see Eq.(\ref{eq:MPeff}), with $M_P$ the reduced Planck mass. We shall discuss at the end of this section the naturalness of this small value in the context of Bjorken's approach discussed before.

Let us recall from our discussion in Sect. \ref{sect:HiggsPotentialSM}  that $\eta\gamma>0$. Therefore, since $\eta<0$ we must also have $\gamma<0$. From (\ref{eq:Geff}) and the last scaling relation in (\ref{eq:power law}) it follows that the effective gravitational coupling is slowly increasing with time:
 \begin{equation}\label{eq:Gt}
 G(t)=G_1\,\left(\frac{t}{t_1}\right)^{-\gamma}\simeq \frac{G_1}{1+\gamma\,\ln\frac{t}{t_1}}\ \ \ \ (\gamma<0, |\gamma|\ll1)\,.
 \end{equation}
 Put another way, Eq.\,(\ref{eq:Gt}) tells us that the gravitational interaction in the current BD-formulation is an asymptotically free theory because $G$ is smaller in the past, which is the epoch when the Hubble rate (with natural dimension of energy) is higher. Equation (\ref{eq:Gt}) can actually be rewritten in a way that makes this fact even more apparent. With the help of the first scaling relation in (\ref{eq:power law}), we can write $G$ directly as a function of $H$:
  \begin{equation}\label{eq:Gt2}
 G(H)\simeq \frac{G_1}{1+\nu\,\ln\frac{H^2}{H^2_1}}\ \ \ \ (\nu>0, |\nu|\ll1)\,,
 \end{equation}
 where we have defined $\nu\equiv-\gamma/2>0$. Clearly, $G(H)$ decreases when $H$ increases, and vice versa. In this form, Eq.\,(\ref{eq:Gt2}) incidentally adopts the same structure as it was found in a class of models with dynamical vacuum and gravitational coupling-- see e.g.\,\cite{JSPReview2013,GRF2015,SolaGomez2015,JSPReview2016} and references therein. In that context, $\nu$ plays the role of the coefficient of the $\beta$-function for the running of these parameters. A theoretical estimate of $\nu$ (hence of $\gamma$) in such framework is that its value should be at most in the ballpark of $10^{-3}$\,\cite{Fossil07}.  This is consistent with the aforementioned limits on the possible time variation of the gravitational constant\,\cite{Uzan2011,Chiba2011}.

Equation (\ref{eq:Higgs mass}) reveals quite evidently another surprise, to wit: the mass of Higgs field changes mildly with time. Most important,  the electroweak VEV itself is depending on the cosmic time, and the variation goes as $v\sim t^{\gamma/2}$, or since $\gamma$ is very small:
\begin{equation}
v^2(t)=\frac{M^2_H(t)}{2\lambda}
\simeq{\frac{2|\eta|\psi_1}{\lambda}}\,{\left(1+\,\gamma\,\ln\frac{t}{t_1}\right)}\,.\label{eq:VeV}
\end{equation}
The  consequences are potentially far reaching: the masses of all the fundamental fermions (quarks and leptons) and gauge bosons, being proportional to the VEV, should change slowly (decrease, since $\gamma<0$) with time. The evolution is logarithmic and therefore very mild, let alone that $|\gamma|\ll1$. This is welcome, of course, since the time evolution of the fundamental ``constants'' should remain moderate enough. For the leading relations, therefore, the log term can be ignored. In the spirit of Bjorken's approach outlined in the previous section, we now use Eq.\,(\ref{eq:gaugelargelog}) to eliminate $\lambda$ in Eq.\,(\ref{eq:VeV}), together with $\psi_1\simeq M_P^2=m_P^2/8\pi$, and we find the approximate result:
\begin{equation}\label{eq:vBjorken}
v^2\simeq \left(\frac{|\eta|} {16\pi^3}\,\ln\frac{m_P^2}{m_{\CC}^2}\right)\,m_P^2\,.
\end{equation}
The above equation is a suggestive form of expressing the EW vacuum expectation value in terms of only parameters of our DB-gravity approach and the CC scale. What is more, Eq.\, (\ref{eq:vBjorken}) appears to be a particular implementation of (\ref{eq:v2lambda1}), in which Bjorken's ratio $r_{Bj}$ is here essentially replaced by a new one defined as
\begin{equation}\label{eq:relationreta}
r_{BD}\equiv{\frac{|\eta|}{16\pi^3}\,\ln{\frac{m_P^2}{m^2_{\CC}}}}\equiv \kappa {|\eta|} \ \ \ \ (\kappa\sim 1)\,.
\end{equation}
This ratio is of the same order of magnitude as Bjorken's ratio and we may call it ``BD-ratio''.
As we can see, it is essentially characterized by the parameter of our extended BD-approach, $\eta$.
It goes without saying that $\kappa$, of order one, cannot be given an accurate numerical value at this point  since the relations (\ref{eq:gaugelargelog}) are true only within order of magnitude. Still, the ratio (\ref{eq:relationreta}), with $|\eta|$ given by (\ref{eq:etaestimate}), is numerically close to $v^2/m_P^2$, and hence Eq.\,(\ref{eq:vBjorken}) is perfectly consistent. For our purposes, $r_{BD}\sim |\eta|$ is a good approximation. In the original Bjorken's formulation, the value $r_{Bj}\sim 10^{-33}$ is given by Eq.\,(\ref{eq:v2lambda}), namely a weighted ratio between the CC scale and the Planck scale -- the weight factor being the large log (\ref{eq:largelog}). Thus the EW scale vanishes if the CC vanishes. In our case we have a similar situation, which is numerically very close, but it does not involve the quotient of the scale $m_{\CC}$ and the Planck mass, but the quotient of the  Higgs mass squared and the reduced Planck mass squared, or the EW scale squared and the usual Planck mass squared.  We can indeed write down the BD ratio in different ways as follows:
\begin{equation}\label{eq:ratiosBD}
 r_{BD}\sim |\eta|\sim\frac{v^2}{m_P^2}\sim \frac{M_H^2}{4\psi_1}\simeq 6.6\times 10^{-34}\sim 10^{-33}\,.
\end{equation}
Because $m_{\CC}\sim k\,M_H^2/M_P\sim 10^2\times10^4/10^{18}\,GeV=10^{-3}$ eV, the above ratio is numerically close to Bjorken's ratio. Notice that in the present BD-framework we also find that the EW scale (and its associated contribution to the vacuum energy) vanishes in an appropriate limit, but in this case it corresponds to the vanishing limit of the dimensionless parameter $\eta\to 0$ existing in the original action (\ref{eq:SMBDaction}). The departure of the BD ratio from zero, according to the measured value of $M_H$, is what allows to recover the EW scale as a tiny fraction of the Planck scale: $\sqrt{r_{BD}}\sim 2.5\times 10^{-17}$.

If Bjorken's attempt at finding a relation between the parameters of the SM and those of gravity is considered a worthy one, we should expect that tiny dimensionless numbers such as $r_{Bj}$ or $r_{BD}\sim \eta$, depending on the formulation, must have a true \textit{raison d'\^etre} which must ultimately be connected to the solution of the CC problem. Let us, however, emphasize that small numbers are not necessarily related to fine tuning. While producing a small scale from the cancelation of two big ones is what we would undoubtedly call real fine tuning (even if all quantities were dimensionless) we, instead, followed Bjorken's road and produced a small (dimensionless) number from a ratio of two widely separated scales. This is, obviously, a very different thing. Such feature could be expected from an attempt at conceiving the CC problem as a hierarchy problem, rather than as a fine-tuning problem, and it may help to find its eventual solution through the theoretical mechanisms connected to the former.

On the phenomenological side, the above scenario with time dependent masses could have important implications, e.g. leading to variations of the proton-to-electron mass ratio $\mu\equiv m_p/m_e$. This ratio has been monitored over the years through different high precision experiments in the lab and in the astrophysical domain\cite{Uzan2011,Chiba2011,PlanckConstants2015,CalmetKeller}. Such feature, if confirmed, could have an impact on a possible very slow variation of the masses of all the nuclei and dark matter particles in the Universe.

The foregoing considerations suggest that the generalized formulation of the BD-gravity under study points to the evolution of all the fundamental constants of Nature. These include the masses, as we have seen above, but in actual fact it includes also the gauge and Yukawa couplings as well, if one is to follow Bjorken's  Eq.\,(\ref{eq:gaugelargelog}) where we recall that in our context the Planck mass varies mildly with time since $G$ does the same, cf.\,Eq.\, (\ref{eq:Gt}).

Overall, the frequently noticed possibility that the so-called fundamental ``constants'' of Nature could actually be slowly dynamical variables \cite{Uzan2011,Chiba2011,CalmetKeller,PlanckConstants2015} might find a possible theoretical explanation in the kind of generalized BD-cosmology that we have put forward here, in which not only $G$ changes with time (as it is usual in ordinary BD-gravity) but also all the particle masses and remaining couplings. Such dynamical picture of the fundamental ``constants'' can also be motivated from alternative approaches, see e.g.  \cite{FritzschSola,FritzschSola2015,Flambaum2015,FritzschNunesSola}.

\section{Brans-Dicke gravity in Grand Unified Theories}\label{sec:BDGUT}
The above ideas can be generalized for theories defined at energy scales higher than in the SM, for example in the GUT framework. As an example, in this section we generalize the action for the Brans-Dicke and Higgs boson scalar fields in the $SU(5)$ context. Following a similar structure as in Eq.\,(\ref{eq:SMBDaction}), the extended form of the action is the following:
\begin{eqnarray}
S=\int d^{4}x\sqrt{-g}\left[\frac{R\psi}{2}-\frac{\omega}{2\psi}\partial_{\mu}\psi \partial^{\mu}\psi+\xi^{\prime} R Tr\left(\Phi^{2}\right)-\frac{1}{2}Tr\left(\nabla\Phi)^2\right) +S^{\prime}_{\mu\nu}Tr\left({\Phi^{-2}{\nabla^{\mu}\Phi\nabla^{\nu}\Phi}}\right)\right.\nonumber\\ +\left.\eta^{\prime} Tr\left(\Phi^{2}\right)\psi+\xi R \Pi^{2}-\frac{1}{2}\left(\nabla\Pi\right)^2 +S_{\mu\nu}\left({\Pi^{-2}}{\nabla^{\mu}\Pi^{\dagger}\nabla^{\nu}\Pi}\right)
+\eta \Pi^{2}\psi -V(\Phi,\Pi)\right]\,. \label{eq:lagrangian-GuT}
\end{eqnarray}
The BD- field is denoted  $\psi$, as before, and behaves as a $SU(5)$ singlet. {The additional scalar field ingredients are, on the one hand, $\Phi$ (the 24-plet in the adjoint representation of $SU(5)$) and on the other $\Pi$ (in the vector representation of the same group). The former is represented by a  $5\times5$ traceless Hermitian Higgs boson matrix, $\Phi$, whereas the latter is the Higgs $5$-plet of the fundamental representation, $\Pi$. As we know the  Higgs doublet of the SM is integrated in the $5$-plet representation of $SU(5)$. The trace in the terms involving $\Phi$ is needed to generate $SU(5)$ invariants.} Some obvious notation used in (\ref{eq:lagrangian-GuT})  is as follows:  $(\nabla\Phi)^2\equiv g^{\mu\nu}\nabla_{\mu}\Phi\nabla_{\nu}\Phi$, $(\nabla\Pi)^2\equiv g^{\mu\nu}\nabla_{\mu}\Pi^{\dagger}\nabla_{\nu}\Pi$, $\Pi^2\equiv \Pi^\dagger\Pi$, which we will use when there is no ambiguity. Similarly, $\Phi^{-2}\equiv\left(\Phi^2\right)^{-1}$ is the inverse matrix of $\Phi^2$, and $\Pi^{-2}=1/ \Pi^\dagger\Pi$.

All the couplings are dimensionless parameters. Those associated to the $\Pi$ multiplet will be denoted without a prime, as in the SM case studied in Sect.\,\ref{set:BDinSM} (recall that the SM Higgs doublet is part of it), whereas the couplings associated to the 24-plet representation of $SU(5)$, i.e. $\Phi$, will be denoted with primes. For example,  $\xi$ and $\xi^{\prime}$ in Eq.\,(\ref{eq:lagrangian-GuT}) are the corresponding non-minimal coupling of the field multiplets $\Pi$ and $\Phi$ to the Ricci scalar $R$; and $\eta$, $\eta^{\prime}$ are the respective coupling constants of $\Pi$ and $\Phi$ to the BD-field $\psi$.
Similarly, the tensor  $S_{\mu\nu}$ that provides the derivative coupling of curvature  to $\Pi$ is identical to the one that was defined previously in the SM case, while $S'_{\mu\nu}=\varsigma' R_{\mu\nu}-\frac{\theta'}{2}g_{\mu\nu}R$ is introduced here, with the same structure as  $S_{\mu\nu}$,  so as to enable the corresponding derivative coupling  of $\Phi$ to curvature.

In addition, by utilizing a unitary matrix $U$ we can diagonalize the Higgs boson Hermitian matrix  $\Phi$ as follows: $\Phi_{d}=U \Phi U^{\dagger}$, or  $\Phi=U^{\dagger} \Phi_d U$, in which $\Phi_{d}$ stands for a diagonal and traceless matrix: $Tr\Phi_{d}=Tr\Phi=0$ owing to the cyclic property of the trace. In the following, we use $\phi$ instead of $\Phi_{d}$ in order to simplify notation. By the unitarity of $U$, it easily follows that  $\Phi^2=U^{\dagger}\phi^2U$ and  $\Phi^{-2}=\left(U^{\dagger} \phi^2 U\right)^{-1}=U^{\dagger}\phi^{-2}U$, where $\phi^{-2}$ is a diagonal matrix made of the reciprocal of the diagonal elements of $\phi^2$. As a result, the part of the kinetic term for $\Phi$ in the action (\ref{eq:lagrangian-GuT}) can be written in a simplified way, fully in terms of the diagonal matrix $\phi$, as follows:
\begin{equation}\label{eq:kineticPhi}
Tr\left({\Phi^{-2}{\nabla^{\mu}\Phi\nabla^{\nu}\Phi}}\right)=
Tr\left({U^{\dagger}\phi^{-2} U{\nabla^{\mu}U^{\dagger}\phi U\nabla^{\nu}U^{\dagger}\phi U}}\right)=
Tr\left({\phi^{-2}{\nabla^{\mu}\phi\nabla^{\nu}\phi}}\right)=\sum_i \frac{\nabla^{\mu}\phi_{ii}\nabla^{\nu}\phi_{ii}}{\phi_{ii}^2}\,,
\end{equation}
where use has been made anew of the cyclic property of the trace, and $\phi_{ii}\, (i=1,2,...,5)$ are the diagonal elements of $\phi$.
The remaining calculations to derive the structure of the Higgs potential follow the same pattern as in the SM case, but are more involved. A summary of the bulky details in this case are presented in Appendix A. The final result is the following:
\begin{eqnarray}
V=-\frac{{2}\eta^{\prime}\psi_1}{\gamma-{2}}\left(\frac{t}{t_{1}}\right)^{\gamma}\,Tr\left(\phi^{2}\right)+\left(6\xi^{\prime} p(p-2)-\frac{1}{2}\right)\frac{(Tr\phi^{2})^{2}}{Tr(\phi_{1}^2)t_{1}^{2}}-45 p^{2}\theta^{\prime}\frac{Tr(\phi^{4})}{Tr(\phi_{1}^4)t_{1}^{4}}\nonumber\\
-\frac{{2}\eta\,\psi_1}{\gamma-{2}}\,\left(\frac{t}{t_{1}}\right)^{\gamma}\Pi^{2}+\left(6\xi p(p-2)-\frac{1}{2}\right)\frac{(\Pi^{2})^{2}}{\Pi_{1}^2 t_{1}^{2}}  -9 p^{2}\theta\frac{(\Pi^{2})^2}{(\Pi_{1}^2)^2 t_{1}^{4}}\ . \label{eq:GUT-final-v}
\end{eqnarray}
Using the above mentioned relation between $\phi^2$ and $\Phi^2$, and the corresponding one between $\phi^4$ and $\Phi^4$, we obtain $Tr\left(\phi^{2}\right)=Tr\left(\Phi^{2}\right)$ and $Tr\left(\phi^{4}\right)=Tr\left(\Phi^{4}\right)$ in Eq.\,(\ref{eq:GUT-final-v}). In this way it is possible to express the final result in terms of the original $24$-plet field $\Phi$, if desired.

Once more the self-consistent construction of the Higgs potential is based on searching for power law solutions of the cosmological equations which can reproduce the standard epochs of the cosmic evolution, see Eq.\,(\ref{eq:GUT-powerlaw}) in the appendix.

We can easily trace some resemblances of the potential (\ref{eq:GUT-final-v}) with the SM case studied in Sect. \ref{set:BDinSM}.  For example, the quadratic term $\sim \phi^2$ there, here is replaced by a linear combination of $\Pi^2$ and  $Tr(\Phi^{2})$, and similarly with the quartic terms. The sign of the coefficient of $Tr(\Phi^{2})$ must be negative in order to trigger the first stage of the spontaneous breaking of the $SU(5)$ gauge symmetry\,\cite{GUT}. This phase transition should occur at very high energies $\sqrt{\langle Tr(\Phi^{2}\rangle}\equiv M_X\sim 10^{16}$ GeV, in which $SU(5)\rightarrow SU(3)_c\times SU(2)_L\times U_Y(1)$.  Clearly, we must have $\eta'<0$ since $|\gamma|\ll1$. Furthermore, in the present context we may conceive that the next SSB is triggered by the VEV of the 5-plet of the Higgs potential (\ref{eq:GUT-final-v}) in the usual way. Recall that the SM Higgs doublet is integrated in $\Pi$ and for this reason one can arrange that the electroweak phase transition $SU(2)_L\times U_Y(1)\to U_{\rm em}(1)$ ensues, provided the coefficient of $\Pi^2$ is negative, thus $\eta<0$ owing to the smallness of $|\gamma|$. We naturally ascribe this phase to the value $p=1/2$ of the scaling solution (\ref{eq:GUT-powerlaw}) since such transition should occur within the radiation dominated epoch. Later on the Universe enters the non-relativistic matter dominated epoch, corresponding to $p=2/3$.

Early on, at energy scales near  $M_X \sim 10^{16}$ GeV, the $SU(5)$ Higgs potential undergoes the aforementioned  SSB in which $SU(5)$ breaks down into the product of the color group and the EW group. But before reaching this stage the Universe is assumed to have undergone an inflationary phase. This one is characterized by values of $p>1$ in our scaling solutions. These values are perfectly allowed and therefore the structure of the potential is such that the scaling solutions can provide an explanation for the different epochs of the cosmic evolution, starting from inflation. {However, in this study we do not intend to describe the connection between the different epochs, and in particular we do not consider the details of inflation since we focus only on the physics at the epoch where the Higgs potential drives the evolution of the different phase transitions indicated above. Furthermore, as already noted, the scaling solutions we are dealing with can not provide the interpolation between the different phases. For this one would need to find a more general class of solutions of the field equations, which is a task that goes beyond the scope of this work.}

A remarkable point, not present in conventional formulations, is the following.  Owing to the fact that the mass terms for $\Phi$ and $\Pi$ in Eq.\,(\ref{eq:GUT-final-v}) evolve with time  ($\gamma$ is small, but non-vanishing) the vacuum expectation value of the two GUT Higgs multiplets $\Phi$ and $\Pi$, specifically the VEV's $\langle \phi\rangle\equiv M_X$ and $\langle\Pi\rangle$, will mildly evolve also with time and hence the particle masses will do the same, similarly as in the SM case.  However, the fact that the $SU(5)$ unification scale is a function of time, $M_X=M_X(t)$, implies that the running couplings (converging at that scale) will be evolving with time as well. The upshot is that we meet once more the kind of situation with time-evolving fundamental constants mentioned in Sect. \ref{sect:FundamentalConstants}. It is interesting to note that in Refs.\,\,\cite{FritzschSola,FritzschSola2015,FritzschNunesSola} it was found an alternative implementation of the time dependence of the running couplings in GUT's, what suggests that this property might be rather generical and could have important implications regarding the possible time evolution of the fundamental constants\,\,\cite{Uzan2011,Chiba2011,CalmetKeller,PlanckConstants2015}.

Finally, {let us remark that the final form of the GUT potential is, of course, not just that given by Eq.\,(\ref{eq:GUT-final-v}). It must be completed by the mixing terms involving the 5-plet and the 24-plet in a renormalizable and gauge invariant way, i.e. terms of the form}
\begin{equation}\label{eq:MixedTerms}
\lambda_1 \Pi^2Tr{\Phi^2}+\lambda_2\Pi^{\dagger} \Phi^2 \Pi\,.
\end{equation}
{These terms are necessary for the renormalizability of the GUT, and also for the  appropriate assignment of the EW scale $v=246$ GeV to the SM doublet part of $\Pi$ while leaving the color triplet part of $\Pi$ very heavy, of order $M_X$. This is, of course, the standard assumption in $SU(5)$\,\cite{GUT}, and we just assume it here as well. The terms (\ref{eq:MixedTerms}) cannot be generated by the scaling method since such procedure can only account for the terms that are explicitly coupled to the BD-field $\psi$. Since, however, this field is treated as a classical background field it cannot account for the terms connected with the renormalizability of the conventional $SU(5)$ part of the theory, which must remain well-defined at the quantum level. As a result the quantum corrections to the potential are of course part of the complete form of the effective potential, which must include the standard logarithmic terms of the Coleman-Weinberg effective potential (cf. any book of QFT, e.g. \cite{GUT}). These quantum effects, however, are genuinely associated to a loopwise expansion of the potential in powers of the Planck constant and are of course pure quantum effects that add up on top of the classical structure of the potential. We have been able to  motivate here only the tree-level or classical structure of the potential. This is nevertheless  remarkable since such initial structure is completely \textit{ad hoc} in the conventional formulation and is also a priori unrelated to gravity, in stark contrast to the present BD-gravity framework.}

{To summarize, both in the SM and the GUT cases the scaling method within the BD-gravity approach is able to motivate in a suggestive way the full structure of those parts of the potential of the scalar field ($\phi$ in the SM, or $\Phi$ and $\Pi$ in $SU(5)$) that is coupled to the BD-field $\psi$, but the final form of the potential needs in general to be  complemented with the extra terms required by the quantum effects and demanded by the renormalizability of the theory, such as the terms (\ref{eq:MixedTerms}), which do not communicate with the classical background field $\psi$, or the terms associated to the loopwise expansion of the effective potential. In the SM case, however (cf. Sect.\ref{set:BDinSM}), the situation was simpler since  only one kind of scalar field was involved and as a result we found that the coupling of the Higgs field with the BD-field is sufficient to unambiguously derive the renormalizable classical structure of the SM Higgs potential without any further requirement. But to this classical structure one has to add of course the terms associated to the higher order loop corrections by the conventional QFT techniques. }

\section{Discussion and conclusions}

In this paper, we have tried to find a more physical \textit{raison d'\^etre} for the origin of the Higgs potential of the standard model (SM) of particle physics. While such potential is instrumental in the SM context in order to furnish masses for the gauge bosons and fermions (using, in the last case, appropriate Yukawa couplings), its structure is entirely motivated by formal quantum field theory (QFT) arguments related to $SU(2)\times U(1)$ gauge invariance and renormalizability. These properties are very important to insure the internal consistency of the SM and its predictability power. Without them, the theoretical status of the SM would be much weaker and it would have been impossible to make accurate and well-defined predictions of electroweak precision observables in higher orders of perturbation theory. The theoretical and practical success of the SM rests to a large extent upon these QFT features supporting the structure of the Higgs potential. However, the fact that we have not been able, yet, despite much effort, to harmonize the existence of the  SM of particle physics within a more general context where the gravitational interactions are also included, undoubtedly is one of the most preoccupying aspects of the structure of the SM. To this fact we have to add the persistent and extremely tough cosmological constant problem, which remains unsolved after 99 years of having the $\CC$-term in Einstein's equations\,\cite{Einstein1917}. Somehow it suggests that our current \textit{Weltanschauung} of the physical laws of the Cosmos is still substantially incomplete. It seems intuitive to us that such momentous theoretical conundrums cannot be solved without first finding an overarching, if only effective,  theory comprising the essential facts of particle physics and gravitation sharing the same basic principles. As this formidable theory seems very difficult to reach, here in the meanwhile we have provided some ``communicative ansatz''  between these two worlds.

Specifically, we have made an attempt to elucidate the physical origin of the Higgs potential (a fundamental cornerstone of the SM of particle physics) within the context of a gravity theory, as this might provide some clue as to the possible connection of particle physics interactions and the gravitational phenomena. Rather than dwelling on the framework of General Relativity, we have motivated it within the Brans-Dicke approach, in which we can remain arbitrarily close to GR by sufficiently increasing the value of the $\omega$-parameter. We have performed our study both for the SM Higgs potential and for the Higgs potential of a typical Grand Unified Theory, such as $SU(5)$, in which the SM can be embedded. In both cases we have started from a generalized BD-action in which the BD-field is coupled to other scalar fields, which at the same time are also coupled non-minimally to curvature both derivatively and non-derivatively. From this intertwined structure with the gravitational action, we have been able to consistently motivate the form of the effective potential for the scalar fields other than the BD-field. The significant outcome is that the sought-for effective potential takes on the Higgs potential form, and the other salient feature is that the vacuum expectation value of the Higgs field evolves slowly with time. As a result we predict a general time evolution of the particle masses and couplings. It is interesting to mention that the possibility of having time evolving masses and couplings can also be motivated from different perspectives related to dynamical models of the vacuum energy\,\cite{FritzschSola,FritzschSola2015}. {In our case, such prediction is, however, limited by the fact that the Higgs potential structure generated in our generalized BD-framework includes, in the case of the SM, four arbitrary parameters in the action (\ref{eq:SMBDaction}). Despite the basic features can be described under the assumption that two of them are equal ($\varsigma=\theta$), three independent parameters are nonetheless needed (apart from the characteristic parameter $\omega$ of the BD theory). Since, however, the usual Higgs potential introduces already two arbitrary parameters, we could say that we need one more additional parameter to motivate its structure from the present gravitational framework. These parameters can in principle be pinned down phenomenologically, as they  determine the mass and self-coupling of the Higgs boson field as well as the coupling of the Higgs to curvature.  }

By combining our BD-gravity approach with old Bjorken's ideas\,\cite{Bjorken2001a,Bjorken2001b,Bjorken2010} on the possible relation between the parameters of the SM and the basic parameters of gravity, we have found a possible explanation for the needed values of the coupling between the BD-field and the SM Higgs field so as to construct a realistic Higgs potential. It turns out that this value is connected to the hierarchy of masses existent between the Planck mass and the mass scale $m_{\CC}\sim 10^{-3}$ eV of the cosmological constant. In this framework, proposed by Bjorken in the aforementioned papers, one finds that the electroweak vacuum energy would be zero if and only if $m_{\CC}=0$. It could explain why the presence of the electroweak vacuum energy might not be a direct contribution to the cosmological constant, and therefore it could make compatible the existence of the Higgs boson with the measured value of the CC.  In this work, building in part upon these interesting Bjorken's ideas, we have considered a more physical motivation for the origin of the Higgs potential of particle physics in the context of a generalized BD-gravity formulation. The essential contribution of the BD-approach in this context is that by establishing a coupling of the BD-field to the Higgs field we can generate most of the structure of the Higgs potential by consistence arguments insuring the existence of the particular cosmological solutions of the field equations characterizing the standard epochs of the cosmic expansion, ranging from the inflationary epoch to the radiation and matter dominated epochs before entering the present dark energy epoch. It could be a possible scenario where to set up some preliminary connection between the particle physics phenomena in the subatomic world and the large scale structure of the Universe, and it might give some new clues for solving the cosmological constant problem. In the meanwhile, a general time evolution of the fundamental constants is predicted, which might act as a smoking gun of the underlying theory.


{\bf Acknowledgments}

\noindent JS has been supported in part
by  FPA2013-46570 (MICINN), Consolider grant CSD2007-00042 (CPAN),
2014-SGR-104 (Generalitat de Catalunya) and MDM-2014-0369 (ICCUB). EK has been supported in part by Bu-Ali-Sina Univ. Hamedan and is thankful to the Ministry of Science, Research and Technology of Iran for financial support; she would also like to thank the Departament de F\'\i sica Qu\`antica i Astrof\'\i sica (formerly Dept. ECM during her visit), Univ. de Barcelona, for support and hospitality during the realization of the initial stages of this work.

\appendix

\section{Higgs potential from BD-gravity in the GUT case}

In this appendix we provide some more details on the self-consistent derivation of the Higgs potential in the GUT case, following similar steps to those we have exposed in detail in Sect.\ref{set:BDinSM} for the SM case.

Let us start by taking the variational derivative of the BD-gravity action (\ref{eq:lagrangian-GuT}) with respect to the metric in the GUT context. We follow similar steps as with the corresponding BD-action in the SM case (cf. Sect. \ref{sect:eqmotionSM}), but we avoid now to write down the general formulae in an arbitrary metric since they are too cumbersome. Expressing the final result directly in terms of the spatially flat FLRW metric, Eq.(\ref{eq:frwmetric}), we arrive at the following result after some algebra:
\begin{eqnarray}
&&{{3}H^{2}\psi}+{{3}H{\dot{\psi}}}-\frac{\omega}{2}\frac{\dot{\psi}^{2}}{\psi}+\eta^{\prime} Tr\left(\phi^{2}\right)\psi-\frac{1}{2}Tr\left(\dot{\phi}^{2}\right)-V(\phi,\Pi)+6\xi^{\prime} H^{2}Tr\left(\phi^{2}\right)+12\xi^{\prime} H Tr\left(\dot{\phi}\phi\right)\nonumber\\
&-&{9}\theta^{\prime} H^{2}Tr\left(\dot{\phi}^{2}\phi^{-2}\right)- 6\left(\theta^{\prime}-\varsigma^{\prime}\right)\dot{H}Tr\left(\dot{\phi}^{2}\phi^{-2}\right)+6\left(\theta^{\prime}-\varsigma^{\prime}\right)H \left[Tr\left(\dot{\phi}\ddot{\phi}\phi^{-2}\right)-Tr\left(\dot{\phi}^{3}\phi^{-3}\right)\right] \nonumber\\
&+&\eta \Pi^{2}\psi-\frac{1}{2}\left(\dot{\Pi}^{2}\right)-{9}\theta H^{2}\left(\dot{\Pi}^{2}\Pi^{-2}\right)+6\xi H\left(H\Pi^{2}+\dot{\Pi}^{\dagger}\Pi+\Pi^{\dagger}\dot{\Pi}\right)- 6\left(\theta-\varsigma\right)\dot{H}\left(\dot{\Pi}^{2}\Pi^{-2}\right) \nonumber \\
&+&3\left(\theta-\varsigma\right)H \left[\sum_{i=1}^{5}\left(\ddot{\overline{\Pi}}_{i}\dot{\Pi}_{i}+\dot{\overline{\Pi}}_{i}\ddot{\Pi}_{i}\right)\Pi^{-2}-
\sum_{i=1}^{5}\left(\dot{\overline{\Pi}}_{i}{\overline{\Pi}_{i}}^{-2}\Pi^{-1}_{i}+\dot{\Pi}_{i}\Pi^{-2}_{i}{\overline{\Pi}_{i}}^{-1}\right)\dot{\Pi}^{2}\right]=0\,.  \label{eq:GUT-EoM-metric}
\end{eqnarray}
As usual the overdot denotes derivative with respect to the cosmic time, and a bar denotes complex conjugation. Following a similar pattern of notation as before,  we have defined $\dot{\Pi}^{2}=\dot{\Pi}^\dagger \dot{\Pi}$.

Now it is turn to take variation with respect to $\phi$, but at this point we must be careful since the components of the diagonal matrix $\phi$ are not independent objects, ergo it is convenient to use Lagrange multipliers to deal with the constraint  $Tr\phi=0$. Performing the variation for each component leads to the field equations for each one of them, namely
\begin{eqnarray}
\left(\ddot{\phi_{ii}}\right)+3 H\left(\dot{\phi_{ii}}\right)-12\xi\left(\dot{H}+ 2 H^{2}\right)\left(\phi_{ii}\right) +18 \left( 2 \theta' -\varsigma' \right) {H}^{3}\left(\dot{\phi_{ii}}\phi^{-2}\right)+\frac{d V(\phi_{ii})}{d\phi_{ii}} \nonumber\\+6 \left(2\theta' -\varsigma'\right) H^{2}\left(\ddot{\phi_{ii}}\phi_{ii}^{-2}-\dot{\phi_{ii}}^{2}\phi_{ii}^{-3}\right)+
6 \left(7 \theta' -5 \varsigma' \right) H \dot{H}\left(\dot{\phi_{ii}}\phi_{ii}^{-2}\right)-{2}\eta^{\prime} \psi\left(\phi_{ii}\right) \nonumber\\+6 \left(\theta' - \varsigma' \right)\ddot{H}\left(\dot{\phi_{ii}}\phi_{ii}^{-2}\right)+
6 \left(\theta' -\varsigma'\right)\dot{H}\left(\ddot{\phi_{ii}}\phi_{ii}^{-2}-\dot{\phi_{ii}}^{2}\phi_{ii}^{-3}\right)-\lambda=0 \,.\label{eq:GUT-EoM-phi}
\end{eqnarray}
Here $\lambda$ is an unknown constant multiplier, and $\phi_{ii}$ denotes the ith-component of the diagonalized $5\times5$ Higgs field matrix $\Phi$, following the procedure explained in Sect. \ref{sec:BDGUT}.

For $\Pi$ and $\Pi^\dagger$, upon taking the variation for each component and then doing a summation over all components, we are led to the field equation for each component $\Pi_i$:
\begin{eqnarray}\label{eq:GUT-EoM-pi}
\frac{1}{2}\left(\ddot{\overline{\Pi}}_{i}+3 H \dot{\overline{\Pi}}_{i}\right)+\frac{d V\left(\Pi_{i}\right)}{d\Pi_{i}}+3\left( \left(2\theta -\varsigma\right) H^{2}+ \left(\theta -\varsigma\right)\dot{H} \right)\left[\frac{\ddot{\overline{\Pi}}_{i}}{\Pi^2}-\frac{ \overline{\Pi}_{i} \dot{\Pi}^2 }{ (\Pi^2)^2}\right]-\eta \overline{\Pi}_{i} \psi \nonumber\\+3\left( \left(\theta - \varsigma \right)\ddot{H} +
3 \left( 2 \theta -\varsigma \right) {H}^{3}+
\left(7 \theta -5 \varsigma \right) H \dot{H}\right)\frac{\dot{\overline{\Pi}}_{i}}{\Pi^2}-6\xi\left(\dot{H}+2H^{2}\right)\overline{\Pi}_{i} = 0\,.
\end{eqnarray}
Likewise, the field equation for the components of $\Pi^\dagger$ will be the same as that for $\Pi$, apart from replacing ${\Pi}_{i}$ with its complex conjugate $\overline{\Pi}_{i}$. Finally, taking the variation of the action with respect to $\psi$, we find:
\begin{equation}
3\dot{H}+6{H}^{2} -  \omega \frac{\ddot{\psi}}{\psi}+ \frac{\omega}{2} \frac{\dot{\psi}^{2}}{{\psi}^{2}}-3H\omega\frac{\dot{\psi}}{\psi}+ \eta Tr\left({\phi} ^{2}\right)+ \eta \Pi ^{2}=0.  \ \ \ \ \ \ \ \ \ \ \label{eq:GUT-EoM-psi}
\end{equation}
As with the SM case, we now seek for a family of power-law (scaling) solutions of the field equations, both for the scalar fields and Hubble parameter:
\begin{equation}
H=\frac{p}{t} \ \ (p>0)\,, \ \ \ \phi_{ii}(t)=(\phi_{ii})_{1}\left(\frac{t}{t_{1}}\right)^{\alpha}, \ \ \ \ \ \Pi_{i}(t)=(\Pi_{i})_{1}\left(\frac{t}{t_{1}}\right)^{\beta}, \ \ \ \
\psi(t)=\psi_{1}\left(\frac{t}{t_{1}}\right)^{\gamma}\,, \label{eq:GUT-powerlaw}
\end{equation}
where  $(\phi_{ii})_{1}$, $\psi_{1}$ and $(\Pi_{i})_{1}$ are the values of correspondence fields at $t=t_{1}$, $p$ is a positive constant and $\alpha$, $\beta$ and $\gamma$ are parameters that we would try to determine in the following by looking for a consistent determination of the potential from the various equations (\ref{eq:GUT-EoM-metric}), (\ref{eq:GUT-EoM-phi}) and (\ref{eq:GUT-EoM-pi}) involving it. We start by substituting the scaling ansatz in Eq. (\ref{eq:GUT-EoM-metric}). This yields a first straightforward form for the potential in terms of the parameters of the scaling ansatz:
\begin{eqnarray}
V\left(\phi,\Pi \right)=\frac{\psi_{1}}{ t_{1}^{\gamma}}\left({3p(p+\gamma)}- \frac{\omega \gamma^{2}}{2}\right)t^{\gamma-{2}}+Tr\left(\phi_{1}^{2}\right)\left(6\xi^{\prime} p\left(p+2\alpha\right)-\frac{1}{2}\alpha^{2}\right)\frac{ t^{2\alpha-2}}{t_{1}^{2\alpha}}+\eta^{\prime} Tr\left(\phi_{1}^{2}\right) \psi_{1}\frac{t^{{2}\alpha+\gamma}} {t_{1}^{{2}\alpha+\gamma}} \nonumber \\ +{\Pi_{1}}^{2}\left(6\xi p(p+2\beta)-\frac{1}{2}\beta^{2}\right)\frac{ t^{2\beta-2}}{t_{1}^{2\beta}}+\eta \Pi_{1}^{2} \psi_{1}\frac{t^{{2}\beta+\gamma}} {t_{1}^{{2}\beta+\gamma}}- 9 p^{2}\left(5\theta^{\prime} \alpha^{2}+\theta \beta^{2} \right) t^{-4}. \nonumber\\ \label{eq:GUT-EoM-metric-1}
\end{eqnarray}
In the previous expression $Tr\phi_{1}^{2}\equiv\sum_{i=1}^5(\phi^2_{ii})_{1}$ is the summation over all diagonal components $(\phi^2_{ii})_{1}$, and similarly $\Pi_{1}^2=\sum_{i=1}^{5}\left(\overline{\Pi}_{i}\right)_{1} \left(\Pi_{i}\right)_{1}$. Note that $Tr\left(\phi_{1}^{2}\phi_{1}^{-2}\right)=5$, but $\Pi_{1}^2\Pi_{1}^{-2}=1$. Inserting next the scaling ansatz (\ref{eq:GUT-powerlaw}) into  the other field equation, Eq. (\ref{eq:GUT-EoM-phi}), we can determine algebraically the derivative ${d V(\phi_{ii})}/{d\phi_{ii}}$, which is some function of time, $f_i(t)$, after we use the power-law solution(\ref{eq:GUT-powerlaw}). But then, of course, we have to integrate $V=\int f_i(t)\dot{\phi}_{ii}dt=\alpha\,((\phi_{ii})_1/t_1)\int f_i(t)\,(t/t_1)^{\alpha-1}\,dt\ $. In this way we are led to an alternative form for the potential:
\begin{eqnarray}
V(\phi_{ii})=-\frac{ \alpha }{\left({2}\alpha-2\right)}\left[\alpha\left(3p+\alpha-1\right)-12\xi^{\prime} p\left(2p-1\right)\right]((\phi_{ii})_{1})^{2} \frac{t^{{2}\alpha-2}}{ t_{1}^{{2}\alpha}}+\frac{{2}\eta^{\prime} \psi_{1}  \alpha}{\left({2}\alpha+\gamma\right)}((\phi_{ii})_{1})^{2}\frac{t^{{2}\alpha+\gamma}} {t_{1}^{{2}\alpha+\gamma}} \nonumber \\ -\frac{9}{2} p\alpha^{2}\sigma^{\prime} t^{-4}+\lambda (\phi_{ii})_{1}\frac{t^{\alpha}}{t_{1}^{\alpha}}\,, \ \ \label{eq:GUT-EoM-phi-1}
\end{eqnarray}
where we have set the integration constant to zero. Here we have defined the combination of parameters  $\sigma^{\prime}=\left[\left(3\theta^{\prime} -2\varsigma^{\prime}\right)p -\left(2\theta^{\prime}-\varsigma^{\prime}\right)p^{2}-\left(\theta^{\prime}-\varsigma^{\prime}\right)\right]$. For $\theta^{\prime}=\varsigma^{\prime}$, it yields  $\sigma^{\prime}=\varsigma^{\prime} p \left(1-p\right)$.

Now, considering  $V(\phi)=\sum\limits_{i=1}^5 V(\phi_{ii})$ and $Tr\phi=0$ (from which the $\lambda$-dependence disappears) we arrive at
\begin{eqnarray}
V(\phi)=-\frac{ \alpha }{\left({2}\alpha-2\right)}\left[\alpha\left(3p+\alpha-1\right)-12\xi^{\prime} p\left(2p-1\right)\right]Tr\left(\phi_{1}^{2}\right) \frac{t^{{2}\alpha-2}}{ t_{1}^{{2}\alpha}}-\frac{45}{2} p\alpha^{2}\sigma^{\prime} t^{-4}\nonumber \\ +\frac{{2}\eta^{\prime} \psi_{1} \alpha}{\left({2}\alpha + \gamma\right)}Tr\left(\phi_{1}^{2}\right)\frac{t^{{2}\alpha+\gamma}} {t_{1}^{{2}\alpha+\gamma}} . \label{eq:GUT-EoM-phi-2}
\end{eqnarray}
Concerning $\Pi$ and $\Pi^{\dagger}$, we determine still another contribution to the potential from   Eq. (\ref{eq:GUT-EoM-pi}), using the power law solution and integrating. Defining  $V(\Pi)=\sum\limits_{i=1}^5 V(\Pi_{i1})+\sum\limits_{i=1}^5 V(\overline{\Pi}_{i1})$, we are led to the explicit form
\begin{eqnarray}
V(\Pi)=-\frac{ \beta  }{\left({2}\beta-2\right)}[\beta \left(3p+\beta-1\right)-12\xi p(2p-1)]\Pi_{1}^{2}\frac{t^{{2}\beta-2}}{ t_{1}^{{2}\beta}}-\frac{9}{2} p\beta^{2}\sigma t^{-4}+\frac{2\eta  \beta\psi_{1}\Pi_{1}^{2}}{\left({2}\beta+\gamma\right)}\frac{t^{{2}\beta+\gamma}} {t_{1}^{{2}\beta+\gamma}}\,, \label{eq:GUT-EoM-pi-1}
\end{eqnarray}
where $\sigma=\left[\left(3\theta -2\varsigma\right)p -\left(2\theta-\varsigma\right)p^{2}-\left(\theta-\varsigma\right)\right]$ which, as before, for $\theta=\varsigma$ we have $\sigma=\varsigma p \left(1-p\right)$ .\\
As for the equation of motion for $\psi$, substituting also (\ref{eq:GUT-powerlaw}) into (\ref{eq:GUT-EoM-psi}) we arrive at the important consistency condition
\begin{equation}
\frac{\eta^{\prime} Tr\left(\phi_{1}^{2}\right) t^{{2}\alpha}}{t_{1}^{2\alpha}}+\frac{\eta \Pi_{1}^{2} t^{{2}\beta}}{t_{1}^{2\beta}}= \left[\omega \gamma \left(\frac{\gamma}{2}+3p-1\right)+{3 p \left(1-{2}p\right)}\right] t^{-2}.  \label{eq:GUT-EoM-psi-1}
\end{equation}
The above equation reveals that in order for both sides to be consistent with each other we must have $\alpha=\beta=-1$, which means that Higgs fields decrease with time, as it should be expected. As a result, Eqs. (\ref{eq:GUT-EoM-phi-2}) and (\ref{eq:GUT-EoM-pi-1}) read as follows:
\begin{eqnarray}
V(\phi)=\left[\left(\frac{3p-2+12\xi^{\prime} p\left(2p-1\right)}{4}\right)Tr\left(\phi_{1}^{2}\right)t_{1}^{2}-\frac{45}{2}p\sigma^{\prime} \right]t^{-4}-\frac{2\eta^{\prime} \psi_{1}}{\gamma-2} Tr\left(\phi_{1}^{2}\right) \frac{t^{\gamma-2}}{ t_{1}^{\gamma-2}}\nonumber\\ \label{eq:v(phi)}
\end{eqnarray}
and
\begin{eqnarray}
V(\Pi)=\left[\left(\frac{3p-2+12\xi p(2p-1)}{4}\right)\Pi_{1}^{2}t_{1}^{2}-\frac{9}{2}p\sigma\right]t^{-4}-\frac{2\eta \psi_{1} \Pi_{1}^{2}}{\gamma-2} \frac{t^{\gamma-2}}{ t_{1}^{\gamma-2}}\,. \label{eq:v(pi)}
\end{eqnarray}
We note that Eqs.\,(\ref{eq:v(phi)}), (\ref{eq:v(pi)}) correspond to the parts of the final potential which depend on the $\phi$ and $\Pi$ fields respectively, while Eq. (\ref{eq:GUT-EoM-metric-1}) is the combination of both fields $\phi$ and $\Pi$. Therefore, consistency of these results requires that $V(\phi,\Pi)=V(\phi)+V(\Pi)$. It follows that the addition of the coefficient for $t^{\gamma-2}$ in Eqs. (\ref{eq:v(phi)}) and (\ref{eq:v(pi)}) must be equal to the coefficient of  $t^{\gamma-2}$ in Eq. (\ref{eq:GUT-EoM-metric-1}). So the following restriction is obtained:
\begin{equation}
\eta^{\prime} Tr(\phi_{1}^{2})t_{1}^{2}+\eta \Pi_{1}^{2}t_{1}^{2}=\frac{\gamma-2}{\gamma}\left(\frac{\omega\gamma^2}{2}-3p(p+\gamma)\right) \label{eq:phi-pi-relation}
\end{equation}
As can be easily recognized, this equation is the counterpart to Eq.\,(\ref{eq:restriction-eta}) in the GUT framework.
Using Eq. (\ref{eq:GUT-EoM-psi-1}) in the previous expression and recalling our previous finding $\alpha=\beta=-1$, we reach the expression
\begin{equation}
\gamma^2\left(1+\omega\right)-\gamma\left(1+p\right)-2p=0\,, \label{eq:GUT-restriction-1}
\end{equation}
which, interestingly enough, is the same constraint obtained in the SM case -- confer Eq. (\ref{eq:restriction-3}). Its solution is given by Eq.\, (\ref{eq:restriction-gammaEXACT}), but in good approximation we retrieve the relation
\begin{equation}
\gamma\simeq -\sqrt{\frac{2p}{\omega}} \label{eq:restriction-gammaGUT}\,,
\end{equation}
in which we keep only the negative sign for $\gamma$. This follows from the generalization of the argument we used for the SM in Sect.\,\ref{sect:HiggsPotentialSM}, namely from the fact that $\eta$ and $\eta'$ are both negative (to enable SSB both in the mass term of the 24-plet and the 5-plet of the Higgs potential in the GUT case) and hence for $p>1/3$ the expression (\ref{eq:phi-pi-relation}) implies (using $|\gamma|\ll 1$) that $\eta\gamma>0$ and $\eta'\gamma>0$, and hence $\gamma<0$.

Similar considerations for the coefficients of $t^{-4}$ in Eqs. (\ref{eq:v(phi)}), (\ref{eq:v(pi)}) and (\ref{eq:GUT-EoM-metric-1}) lead to the relation
\begin{equation}
(1+12\xi^{\prime})Tr(\phi_{1}^{2})t_{1}^{2}+(1+12\xi)\Pi_{1}^{2}t_{1}^{2}=30\sigma^{\prime}+6\sigma-12p(5\theta^{\prime}+\theta)\,.
\end{equation}
Once we have analyzed the various forms of the potential and determined the values of the parameters in the scaling ansatz (\ref{eq:GUT-powerlaw}) such that we can obtain a consistent solution, the final form of the potential can be made explicit. Thus, for instance,  since $\alpha=\beta=-1$ we may take the trace to the quadratic and quartic powers of the second relation in (\ref{eq:GUT-powerlaw}) and sum over the third  such that the form of the potential in  Eq. (\ref{eq:GUT-EoM-metric-1}) can be written precisely as the GUT Higgs potential indicated in the text, cf. Eq.\, (\ref{eq:GUT-final-v}).  We can see indeed from Eq.\.(\ref{eq:GUT-EoM-psi-1}) that the $\eta$ and $\eta'$ couplings of the BD field $\psi$ to the two kinds of Higgs multiplets in the action (\ref{eq:lagrangian-GuT}) are essential to fix the unique values $\alpha=\beta=-1$ in the scaling relations.

Finally, the mixing couplings between $\Phi$ and $\Pi$ mentioned in Sect.\,\ref{sec:BDGUT}, cf. Eq.\,(\ref{eq:MixedTerms}), which are necessary to complete the structure of the potential, can be invoked on renormalization grounds. Notice that, numerically, $\eta$ and $\eta'$ are very different parameters since they belong to two very different SSB scales. As can be seen from  Eq.\, (\ref{eq:GUT-final-v}), the value of $\eta'$ need not be so small as $\eta$, we typically expect  $\eta'\sim M_X^2/M_P^2\sim 10^{-4}$, where $M_X\sim 10^{16}$ GeV defines the characteristic GUT scale. It defines also the order of magnitude of the mass of the 24-plet of Higgs bosons.

Let us note that, in the GUT case, we may conceive more general values for $p$ in the scaling solutions (\ref{eq:GUT-powerlaw}). In the SM context, in which we naturally placed the electroweak transition in the radiation epoch, we naturally adopted $p=1/2$. Here, however, we can have a wider range of possibilities. In fact, the relation (\ref{eq:restriction-gammaGUT}) can also be applied for $p>1$, leading to an inflationary solution, which is welcome in the early Universe described by a GUT.
Recall from Eq.\,(\ref{eq:EoS}) that $p>1$ indeed implies an EoS value $w<-1/3$ characteristic of an inflationary phase.
After inflation has taken place for some value $p>1$, the Universe evolves into the radiation phase and again we can seek for asymptotic solutions of the power law form (\ref{eq:GUT-powerlaw}) for values $p\simeq 1/2$. In this case we would recover the solutions found in Sect. \ref{sect:HiggsPotentialSM} for the SM since the Higgs doublet of the SM is part of the  5-plet $\Pi$  of SU(5) and at low energies the quintuplet piece of the GUT Higgs potential (\ref{eq:GUT-final-v}) boils down to the SM form (\ref{eq:VHiggs}), where $\eta$ is a common parameter. Subsequently the Universe evolves into the dark matter epoch ($p=2/3$) and eventually in the dark energy epoch. But the various stages of the cosmic evolution can be linked only through more general solutions of the cosmological equations, certainly more complicated that the power-law solutions that we have considered here.

\end{document}